\documentclass{article}
\usepackage{PRIMEarxiv}

\usepackage[english]{babel}
\usepackage[T1]{fontenc}
\usepackage[utf8]{inputenc} 
\usepackage{algorithm}
\usepackage{algpseudocode}
\usepackage{amsfonts}       
\usepackage{amsmath}
\usepackage{amssymb}
\usepackage{amsthm}
\usepackage{bm}
\usepackage[table]{xcolor}
\usepackage{booktabs, tabularx}
\usepackage{comment}
\usepackage{fancyhdr}       
\usepackage{graphicx}       
\usepackage{hyperref}
\usepackage{lineno}
\usepackage{lipsum}
\usepackage{microtype}      
\usepackage{multirow}
\usepackage[numbers, sort&compress, comma, square]{natbib}\usepackage{nicefrac}       
\usepackage{subfig}
\usepackage{tikz}
\usepackage{url}            
\usepackage{xspace}
\graphicspath{{media/}}     
\newcommand{\X}{\mathcal{X}\xspace}
\newcommand{\Y}{\mathcal{Y}}
\newcommand{\x}{\mathbf{X}}
\newcommand{\z}{\mathbf{Z}}

\newcommand{\y}{y}

\newcommand{\COV}{\mathbb{C}\text{ov}} 
\newcommand{\bbR}{\mathbb{R}} 

\newcommand{\cM}{\mathcal{M}} 
\newcommand{\Sn}{\mathcal{S}^n}
\newcommand{\SnP}{\mathcal{S}_{++}^n}
\newcommand{\Si}{\mathcal{S}}
\newcommand{\method}{Phase-SPDNet }

\newcommand{\methodmath}{\mathrm{Phase-SPDNet} }

\title{Geometric Neural Network based on Phase Space for BCI-EEG decoding}

\author{Igor Carrara$^{1}$\thanks{Bruno and Igor are joint first authors.} \quad Bruno Aristimunha$^{2, 3, *}$ \\ \textbf{Marie-Constance Corsi}$^{4}$ \quad \textbf{Raphael Y. de Camargo}$^{3}$ \quad \textbf{Sylvain Chevallier}$^{2}$ \quad \textbf{Th\'eodore Papadopoulo}$^{1}$  \\
\\
\small $^1$ Université Côte d'Azur, Inria d'Université Côte d'Azur, Sophia Antipolis, France, \\
\small $^2$ Université Paris-Saclay, Inria TAU, LISN-CNRS, France, \\
\small $^3$ Universidade Federal do ABC, Santo André, Brazil,\\
\small $^4$ ARAMIS, Inria, Paris Brain Institute, Paris, France. \\
\small Correspond author: Bruno Aristimunha, e-mail: b.aristimunha@gmail.com \\
}

\begin{document}

\maketitle

\begin{abstract}
\textbf{Objective:} 
The integration of Deep Learning (DL) algorithms on brain signal analysis is still in its nascent stages compared to their success in fields like Computer Vision. This is particularly true for BCI, where the brain activity is decoded to control external devices without requiring muscle control.
Electroencephalography (EEG) is a widely adopted choice for designing BCI systems due to its non-invasive and cost-effective nature and excellent temporal resolution. Still, it comes at the expense of limited training data, poor signal-to-noise, and a large variability across and within-subject recordings. 
Finally, setting up a BCI system with many electrodes takes a long time, hindering the widespread adoption of reliable DL architectures in BCIs outside research laboratories. To improve adoption, we need to improve user comfort using, for instance, reliable algorithms that operate with few electrodes. 
\textbf{Approach:} Our research aims to develop a DL algorithm that delivers effective results with a limited number of electrodes. Taking advantage of the Augmented Covariance Method and the framework of SPDNet, we propose the \method{} architecture and analyze its performance and the interpretability of the results. The evaluation is conducted on 5-fold cross-validation, using only three electrodes positioned above the Motor Cortex. The methodology was tested on nearly 100 subjects from several open-source datasets using the Mother Of All BCI Benchmark (MOABB) framework. 
\textbf{Main results:} The results of our \method{} demonstrate that the augmented approach combined with the SPDNet significantly outperforms all the current state-of-the-art DL architecture in MI decoding. 
\textbf{Significance:} This new architecture is explainable and with a low number of trainable parameters.

\end{abstract}


\section{Introduction}
Brain-computer interface (BCI) technology allows direct communication between a user's brain activity and external devices. Originally designed to help people with disabilities~\cite{daly2008brain}, its uses are expanding to other fields, such as rehabilitation and virtual reality~\cite{wen2021combining, ctrl2024generic}.
Different signals acquired from brain activity can be used for such a technology, but Electroencephalography (EEG) is a widely adopted choice, as it is a non-invasive, portable, and inexpensive methodology with a very good temporal resolution. Motor imagery (MI) tasks are largely investigated among the brain activities considered when designing a BCI. As the subject is asked to mentally execute a movement without actually performing it, it provides an asynchronous and internal control scheme with no requirement for muscle capability.

The application of Deep Learning (DL) algorithms has garnered significant attention over several domains, ranging from natural language process~\cite{deng2022benefits} to protein structure prediction~\cite{jumper2021highly}. The field of EEG MI classification is no exception. However, DL algorithms have not yet established themselves in the BCI field as they have in other fields, such as Computer Vision. This is due to several problems: limited data availability, low signal-to-noise ratio in EEG signals, subject variability due to anatomical differences between individuals, and to session variability due to deviations in electrode placement~\cite{roy2019deep}. 

Unlike other fields where DL algorithms thrive on extensive data, in EEG applications, the emphasis is on enhancing user comfort, resulting in smaller datasets with few trials and fewer electrodes. Large-scale EEG systems with numerous electrodes not only require extended calibration periods, causing user tiredness but also introduce complexity, potentially leading to increased error rates and heightened computational demands. Additionally, the increased cost associated with extensive electrode setups deters the widespread development, deployment, and accessibility of BCI technology, particularly for the patients who could benefit the most from it.

This research focuses on developing a novel DL architecture, \method, that outperforms the state-of-the-art classification when using a limited number of electrodes. Building on the Augmented Covariance Method (ACM)~\cite{carrara2024classification} that is an extension of the spatial covariance and the Symmetric Positive Definite (SPD) Neural Network -SPDNet~\cite{huang2017riemannian}, we study the \method{}  impact on the performance when using a reduced number of electrodes. We also conducted a comprehensive analysis of the model size and its explainability with respect to the standard SPDNet. 

Spatial covariance, however, is not the only one that can be extracted from the EEG signal. Another potential candidate is coherence, which provides an alternative perspective on EEG signal characteristics. Historically, coherence has proven to be an inherently unstable feature, which is challenging to compute accurately, resulting in less robust results with respect to the spatial covariance. 
However, the use of information from both covariance and coherence has been shown to increase performance as we are considering an estimator of the interactions between brain areas/electrodes~\cite{fucone:2022}. We are thus interested in studying coherence in combination with SPDNet and especially in using the ACM methodology adapted for the coherence matrix.

Our methodology is tested through a 5-fold cross-validation evaluation
(\emph{Within-Session}), using only three electrodes strategically positioned above the Motor Cortex. To validate our algorithm, we test our approach on almost 100 subjects from openly available datasets using the Mother Of All BCI Benchmark \href{http://moabb.neurotechx.com/docs/index.html}{(MOABB)} framework~\cite{aristimunha-carrara-etal:23}. This research not only contributes to the advancement of EEG MI classification but also emphasizes the importance of developing efficient, user-friendly algorithms in the broader context of BCI technology. Additionally, our study places a strong emphasis on ensuring that our method is both reproducible and interpretable.

The article is organized as follows: In ~\autoref{sec:related}, we examine the current state-of-the-art in Deep Learning (DL) EEG decoding, emphasizing distinctions from our approach. \autoref{sec:Material and Methods} provides an overview of the theoretical foundations of our model and the considered datasets. The obtained results from the Within-session evaluation are presented in ~\autoref{sec:Results}. Subsequently, \autoref{sec:Discussion} focuses on the method's impact and current limitations, specifically focusing on its explainability. Finally, \autoref{sec:Conclusion} summarizes the findings of our study.

\section{Related Work}
\label{sec:related}
\subsection{Machine Learning for EEG Decoding}

In the domain of EEG decoding~\cite{casson2019wearable, King:2020}, translating brain activity into meaningful data has become increasingly dependent on machine learning (ML) methods~\cite{jayaram2018moabb, schirrmeister2017, roy2019deep, cedric2022benchmark, fucone:2022,  biomarkers:2023,braindecoding:2024, ouahidi2023strong, wimpff2024eeg, ouahidi2024unsupervised}. However, here, we diverge by focusing on deep learning techniques instead of a broad ML spectrum, moving beyond mere method comparison to introduce a novel EEG decoding approach.

\citet{schirrmeister2017} demonstrated that Deep Learning (DL) approaches, specifically \emph{ShallowNet} and \emph{DeepNet} can perform on par with conventional machine learning in decoding raw EEG data and can be trained end-to-end, eliminating several feature extraction steps. The application of DL algorithms enhances the generalization capability, enabling it to handle the inherent variability present in EEG signals effectively. This applicability of DL for EEG data is further corroborated by~\cite{lawhern2018eegnet, ingolfsson2020eegtcnet, salami2022eeg, chen2022toward, cedric2022benchmark}. Despite their effectiveness, these DL approaches demand extensive parameter tuning, high power consumption, and large datasets for thorough evaluation. In contrast, our study proposes a streamlined approach, leveraging Riemannian geometry and functional connectivity derivatives, which are less demanding in terms of parameters, enhancing both efficiency and scalability. The efficiency and scalability stems from the geometric transformation these methods apply, which reduces the temporal signal's dimensionality from channel and time dimensions to channel-by-channel dimensions.

\subsection{Deep Riemannian Networks for EEG Decoding}
Incorporating non-euclidean geometry, especially the Riemannian manifold, into EEG decoding has significantly advanced the field~\citep{Huang_Van_Gool_2017, Suh_Kim_2021, pan2022matt, kobler2022spd, Wang_2022_ACCV, Ju2022TensorCSPNetAN, discriminative2023, Lu2023LGLBCIAL, graphSPD2023, USPDNET:2023, math11071570, funcional2023, Wilson2022DeepRN}. A common practice is to compute spatial covariance matrices that capture signal features with the structure of symmetric positive definite (SPD) matrices. These matrices not only enhance signal information regarding topology and amplitude but also offer increased robustness to outliers and noise with Riemannian geometry, maintaining invariance under linear transformations~\cite{pan2022matt, Lu2023LGLBCIAL}. 

\citet{Huang_Van_Gool_2017} work introduced \emph{SPDNet}, a neural network that operates on SPD manifolds. This foundation was used by~\citet{spdnet_eeg:2020}, who applied SPDNet in bio-signal classification, enhancing transfer learning. Building on this, \citet{NEURIPS2019_6e69ebbf} adapted batch normalization for the SPD manifold, and \citet{kobler2022spd} further refined the batch normalization component for EEG Decoding. \citet{pan2022matt} added to this perspective by proposing attention components for the category of the SPDNet neural networks. Follow-up efforts included using residual layers~\cite{Wang_2022_ACCV, USPDNET:2023}, filter bank inputs~\cite{Suh_Kim_2021, Ju2022TensorCSPNetAN, graphSPD2023}, mixing traditional convolution by channels~\cite{Wilson2022DeepRN}, and constraining diffusion models~\cite{li2023spd, fishman2023diffusion}. Contrary to our approach, these methods rely on adapting existing neural network components for the SPD manifold or preserving the dimensionality assumptions, while our method combines the dimensionality expansion of the phase space reconstruction on the SPD manifold.

\subsection{Geometry Transformation in EEG data}
Geometry Transformation (GT)~\cite{roy2019deep, cedric2022benchmark, EEGManyPipelinesPreProcessing:2020, carrara2024classification} in EEG analysis involves standard pre-processing transformations like resampling, band-passing, and filtering, which have been proven to enhance model performance and biomarker discovery~\cite{mne:2013, Jas2016AutorejectAA, ablin2018faster, lotte2018review, biomarkers:2023}. Data augmentation, a regularization geometry transformation, has been acknowledged for its role in enhancing brain decoding in EEG, with its benefits varying depending on the task~\cite{rommel2022cadda, cedric2022benchmark}. While these transformations can improve the learning processing, these steps usually do not consider the non-stationary dynamics properties of the biological signal. Our approach broadens this scope by incorporating transformations that consider phase reconstruction in terms of its components of nonlinear dynamics.

In the context of delay embedding phase geometry transformation, \citet{chen2021high} demonstrated the effectiveness of phase embedding in neural networks for one-dimensional bio-signal reconstruction, but without addressing the non-euclidean nature of the signal. Our methodology, in contrast, considers the non-euclidean geometry, thereby enriching the learning process. The studies by \citet{carrara2024classification} and \citet{ZHOU2024105572} align with our approach in their use of phase delays and (cross-)covariance matrices for EEG decoding. Our work, on the other hand, fully leverages the capabilities of the SPD matrix in Riemannian Neural Networks, allowing for more efficient, interpretable layers and achieving superior performance within challenging scenarios with a reduced number of channels.

\subsection{Functional Connectivity for EEG Decoding}
Beyond the framework of processing the raw brain signal data, researchers in the neuroscience community have been interested in identifying the relationship between brain areas using Functional Connectivity (FC) features \cite{cliff2023unifying,fucone:2022}. 
These features demonstrate that electrode interactions are effective for BCI tasks \cite{fucone:2022, cliff2023unifying, nolte2004identifying, pascual2007instantaneous, barachant2011multiclass, Leeuwis2021functional, li2023coherence, de2023reproducibility, chiarion2023connectivity}. 
Most EEG decoding studies using FC employ brain interaction estimators to construct adjacency matrices for each trial, such as covariance, correlation, and coherence, often combined with dimension reduction techniques. 
We refer to \cite{cliff2023unifying} for an extensive discussion on FC estimators for brain signals. 

\section{Material and Methods}
\label{sec:Material and Methods}
In this section, we briefly recall the EEG decoding problem, the Phase Space Reconstruction - PSR transformation, the Riemannian manifold properties, the SPDNet components and the datasets considered.

\subsection{EEG Decoding}
We consider EEG signals as real-valued matrices. Let us denote each bio-signal window captured during a cognitive task (called an epoch) by $\mathbf{X}_i \in \X = \mathbb{R}^{C \times T}$. Here, $C$ represents the number of channels (electrodes), $T$ is the number of time points in the window, and $i$ indexes the trial $i = 1\ldots N$, where $N$ is the total number of trials. Each epoch corresponds to a specific cognitive task denoted by $\y_i \in \Y$.

EEG decoding aims to construct a function $f: \X \rightarrow \Y$, which effectively maps each trial $\x_{i}$ to its corresponding label $\y_{i}$. We can construct a neural network function represented as $f_{\theta}:\mathcal{X}\rightarrow \mathcal{Y}$, composed of sequence of $l$ functions, formulated as $f_{\theta}=f_{\theta_l} \circ \dots \circ f_{\theta_2} \circ f_{\theta_1}$. In this structure, $\theta_i$ indicates the parameters describing each layer $i$ of the network and $\theta$ is the concatenation of all the parameters $\theta_i$. 

\subsection{Phase space reconstruction for the EEG signal}
When using a reduced number of electrodes, the captured signals will contain only part of the real dynamics of the brain. It is possible to recover part of this information using time delay embeddings based on the Takens's theorem~\cite{takens1981detecting}. This methodology allows the reconstruction of the dynamical system in an alternative space, different from the sensors one, but containing the same dynamical information as the brain. 

\begin{figure}[!ht]
    \centering
    \includegraphics[width=\linewidth]{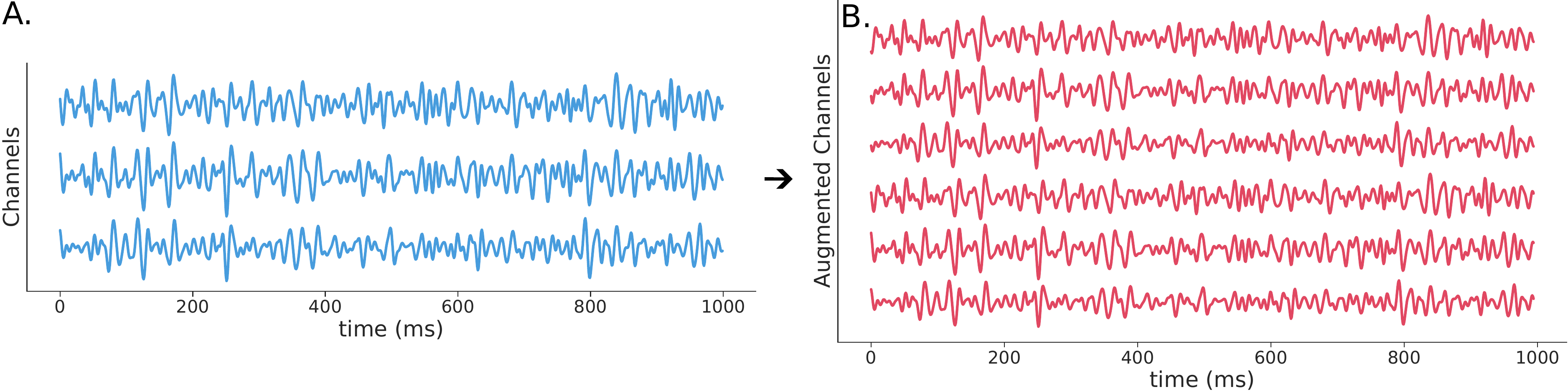}
    \caption{Representation of the augmented procedure. The original signal comprises three electrodes, represented by the plot in blue, while the red plot represents the augmented dataset using embedding parameters $\tau = 5$ and $\psi=2$.}
    \label{fig:augmentation}
\end{figure}

Our method applies Takens' theorem over the minimally pre-processed epoch-cropped time series, which allows for the understanding of a system's multi-variable dynamics through an embedding in a $\psi$-dimensional space by employing a phase embedding transformation as the initial function in our neural network~\cite{takens1981detecting, takens1993, noakes1991takens}. This is made by constructing a phase space using a delay vector constructed from the original signal, thereby enabling the reconstruction of the system dynamics from these observations.
Following the approach proposed by \citet{carrara2024classification}, it is possible to define a delay function $d_p(\mathbf{\x}) = \x[t+p\tau, t+T-(\psi-p)\tau]$ for $p \in [0, \psi]$. This function allows the concatenation of measures on a sliding window over the observable in a delay vector defined as:
\begin{equation}
    f_{\mathrm{delay}}(\x) = [d_0(\mathbf{X}), \ldots, d_{\psi}(\mathbf{X})]^\top
\end{equation}
where $\psi$ represents the embedding dimension, which dictates the order of magnitude of the phase space, and $\tau$ is the embedding delay~\cite{packard1980geometry,embedding_definition:2023}.
As a result, $f_{\mathrm{delay}}(\x)$ embodies an embedding of the original phase space into a higher-dimensional space, enabling a detailed examination of the system's dynamics. This formulation considers the EEG signal as produced by a nonlinear dynamical system. Concerning the dimension, if the original epoch have dimension $\mathbf{X}_i \in \mathbb{R}^{C \times T}$ the matrix $f_{\mathrm{delay}}(\x) \in \mathbb{R}^{C \psi \times (T-\psi \tau)}$. This function is applied on each window signal $\x_{i} \in \X$. An illustration of the process, with $\tau = 5$ and $\psi=2$ is present in ~\autoref{fig:augmentation}.

\subsection{Selection of Phase and Delay}
Selecting the optimal $\tau$ and $\psi$ values can be extremely time-consuming~\cite{carrara2024classification, embedding_definition:2023}. 
To ensure the algorithm operates efficiently, we adopted two distinct methodologies: one grounded in non-linear dynamical systems theory and the other employing a optimized search strategy.

\subsubsection{MDOP}
One first approach to overcome this difficulty was to use the Maximizing Derivatives On Projection (MDOP) method~\cite{nichkawde2013optimal, carrara2024classification} for the multi-epoch context of EEG Decoding, shown in Algorithm \ref{alg:mdop_eeg}. The algorithm relies on the function $mdop\_embedding$ from the package $\mathrm{DelayEmbeddings}$ to implement the MDOP procedure. The function aims to create a nonuniform embedding that generates a vector of different lags $\boldsymbol{\tau}_{i}$.

\begin{algorithm}[!ht]
\caption{MDOP for EEG Decoding}\label{alg:mdop_eeg}
\begin{algorithmic}[2]
\State \textbf{using} DelayEmbeddings: mdop\_embedding
\Function{MDOP for epochs}{$X$}
\State $\tau = 0$
\State $\psi = 0$
\For{$i = 1:\#\text{epochs}$}
    \State $\boldsymbol{\tau}_{i} = \text{mdop\_embedding}(X_{i})$
    \State $\tau = \tau + \lfloor \overline{\boldsymbol{\tau}_{i}}\rfloor$ 
    \State $\psi = \psi + |\text{len} (\boldsymbol{\tau}_{i})| $
\EndFor
\State $\overline{\tau}_{Epoch} =  \lfloor \frac{\tau}{|\#\text{epochs}|} \rfloor$  
\State $\overline{\psi}_{Epoch} =  \lfloor \frac{\psi}{|\#\text{epochs}|} \rfloor$\\
\Return $\overline{\tau}_{Epoch}, \overline{\psi}_{Epoch}$
\EndFunction
\end{algorithmic}
\end{algorithm}

The MDOP algorithm emerged as a novel geometric method for analyzing dynamical systems, as opposed to conventional statistical and information-theoretic methods, such as mutual information and continuity statistics. Central to MDOP is optimizing an embedding to ensure the reconstructed attractor is as expanded as possible while simultaneously reducing redundancy among the delay components. The algorithm employs a recursive approach, wherein each step of the embedding cycle identifies the lag $\tau$ that yields the highest beta statistics. This chosen lag is then used in the subsequent reconstruction phase. This iterative process continues until the algorithm obtains satisfying embedding dimensions $\psi$, selected using the method of false nearest neighbors.

\subsubsection{Searching for better phase and delay parameters based on Optuna}

For hyper-parameter Optimization of the phase and delay, we employ the random search implemented in \textsc{Optuna} version $0.17.0$ \cite{akiba2019optuna}.
We assume a categorical distribution for the variables, $\psi \in [1, 10]$ and $\tau \in [1, 10]$, considering the principles of independence variables found in \cite{carrara2024classification}.
We restricted the search time for each iteration to 15 minutes to ensure a fair comparison between the models. 
The search for the \textsc{Optuna} used the Bayesian Tree-structured Parzen Estimator (TPE) Sampler, as described in \cite{watanabe2023tree}.
To prevent data leakage, we incorporated the hyper-parameter search into a nested cross-validation framework~\cite{cawley2010over}, integrated into the current \textsc{MOABB} version 1.1.

\subsection{Riemannian Manifold}
Symmetric Positive Definite (SPD) matrices have begun to play a key role in several applications, ranging from brain imaging to Computer Vision~\cite{le2001diffusion, weickert2005visualization}. In particular, the classification approach of SPD matrices based on the Riemannian distance algorithm is the current state-of-the-art in the BCI-MI classification~\cite{barachant2010riemannian, lotte2018review}.

Let's define $\cM_n$ the space of real square matrix and $\Sn$ the space of symmetric matrix, where $\Sn = \{\Sn \in \cM_n \mid \mathbf{S}^{\top}=\mathbf{S}\}$. It is now possible to define the space of Symmetric Positive Definite (SPD) matrices as
\begin{equation}
    \SnP=\{S \in \Sn \mid \bm{x}^{\top} \bm{S} \bm{x} > 0  \; \; \; \;      \forall \bm{x} \in \bbR^{n}\}
\end{equation}
This formulation allows us to represent matrices belonging to $\SnP$ as points on a Riemann manifold with a dimension of $n(n+1)/2$. The space of SPD matrices forms a manifold with negative curvature~\cite{forstner2003metric, moakher2005differential}, so Euclidean geometry concepts do not apply.

It is possible to define several distances between two SPD matrices $\Si_1$ and $\Si_2$, generally depending on the length of the geodesic connecting $\Si_1$ and $\Si_2$ on the Riemann manifold. Often, the affine-invariant metric is used in the context of BCI. We proceed, however, to give a mathematical formulation based on the Log-Euclidean metric~\cite{arsigny2007geometric}. This formulation reduces the computational burden associated with the affine-invariant framework while preserving robust theoretical properties~\cite{arsigny2007geometric}.

The space of Symmetric Positive Definite (SPD) matrices can be endowed with a Lie group structure. For a comprehensive understanding, please refer to~\cite{arsigny2007geometric}. It becomes possible to define a distance metric between two SPD matrices $\Si_1$ and $\Si_2$ as the bi-variate metric on the Lie group of SPD matrices
\begin{equation}
    d(\Si_1, \Si_2) = ||\log(\Si_2)-\log(\Si_1)||
\end{equation}
where $|| \quad ||$ is the norm associated with the metric, and $\log$ is the matrix logarithm. The Log-Euclidean metric on the Lie group of SPD matrices corresponds to an Euclidean metric within the logarithmic domain of the SPD matrices.

Distances, geodesics, and Riemannian means exhibit a more straightforward formulation within the Log-Euclidean metric than the affine-invariant case, maintaining comparable invariance properties. However, this simplification comes at the cost of more intricate formulations for exponential and logarithmic mapping. The mapping of $\Si_2$ respect to $\Si_1$ is defined as
\begin{equation}
    \begin{array}{l}
\log _{\Si_1}\left(\Si_2\right)=D_{\log \left(\Si_1\right)} \exp \cdot\left(\log \left(\Si_2\right)-\log \left(\Si_1\right)\right), \\
\exp _{\Si_1}(\Si_2)=\exp \left(\log \left(\Si_1\right)+D_{\Si_1} \log \cdot \Si_2\right) .
\end{array}
\label{eq:log}
\end{equation}

where $D_{\log \left(\Si_1\right)} \exp$ represent the differential at point $\Si_1$ of the exponential function and similarly for $(D_{\log \left(\Si_1\right)} \exp)^{-1}=D_{\Si_1} \log$. This formulation can be simplified if the reference matrix for the mapping is the identity matrix. \autoref{eq:log} presents two quantities, the exponential and logarithmic maps, which are used to transition from Euclidean space to the Riemannian surface. 

In BCI, the spatial covariance is estimated from the pre-processed EEG signal $\x \in R^{C \times T}$. Several estimators can be used to estimate covariance \cite{pyriemann}, but the most popular is the sample covariance matrix
\begin{equation}
    f_{\COV}(\x_{i}) = \frac{1}{T-1}\sum_{i=1}^T \x_i \x_i^\top
\end{equation}

\subsubsection{Symmetric Positive Definite Neural Networks Components}
In the context of geometry neural networks, \citet{huang2017riemannian} proposed the neural network SPDNet with three SPD layers:

\paragraph{\textbf{BiMap Layer:}} The bi-linear mapping level aims at creating more compact and discriminating SPD matrices just like the convolutional layer in a standard DL network, with the complication, however, that SPD matrices live in a Riemannian space. We can describe the equation of the BiMap layer as 
\begin{equation}
    f_{\mathrm{BiMap}}(\z_{k-1}) = \z_k = \textbf{W}_k \z_{k-1} \textbf{W}_k^{\top}\;,
    \label{Bimap}
\end{equation}
where $\z_{k-1} \in \mathcal{S}_{d_{k-1}}^{++}$ and $\textbf{W}$ is, for this layer, the learnable parameter. The matrix $ \textbf{W}_k$ is a full rank matrix in order to guarantee the output is an SPD matrix. Such matrices belong to a noncompact Stiefel manifold making the optimization problem impossible due to the absence of an upper bound on the distance function. To address this, we impose an additional constraint requiring $ \textbf{W}_k$ to be orthogonal. This places the weight matrix on a compact Stiefel manifold, making the optimization problem solvable~\cite{huang2017riemannian}.

\paragraph{\textbf{ReEig Layer:}}
This layer introduces a non-linearity with a similar approach to a Rectified Linear Unit (ReLU) level. In practice, it rectifies SPD matrices by thresholding small eigenvalues to  $\varepsilon$.
\begin{equation}
    f_{\mathrm{ReEig}}(\z_{k-1}) = \z_{k} = \textbf{U}_{k-1}\max(\varepsilon I, \Sigma_{k-1})\textbf{U}_{k-1}^{\top}\;,
\end{equation}
where $\textbf{U}_{k-1}$ and $\Sigma_{k-1}$ are not learned but obtained using eigenvalue decomposition of the previous layer, $\z_{k-1}=U_{k-1}\Sigma_{k-1}U_{k-1}^T$. This layer has no trainable parameters.

\paragraph{\textbf{LogEig:}} 
This layer aims to transport the SPD matrix obtained from the previous layers from a Riemannian space to an Euclidean one using the Log-Euclidean metric~\cite{arsigny2007geometric}. This is formally the expression (\ref{eq:log}) considering the mapping with respect to the identity, so formally, the layer is defined as
\begin{equation}
    f_{\mathrm{LogEig}}(\z_{k-1}) = \z_k = \log(\z_{k-1}) =\textbf{U}_{k-1}\log(\Sigma_{k-1})\textbf{U}_{k-1}^{\top}
\end{equation}
where again, the $\textbf{U}_{k-1}$ and $\Sigma_{k-1}$ are not learned but obtained using eigenvalue decomposition.This layer has no trainable parameters. This logarithmic mapping is applied using the Identity as a reference matrix. 
Once this layer is applied, we can use the classical deep learning method in the Euclidean space as a Multi-Layer Perceptron.

\subsection{Phase-SPDNet}
Given the framework, we define our neural network, denoted as \( f_{\methodmath} \), as a composition of sequential transformations:
\begin{equation}
    f_{\methodmath} = f_{\mathrm{MLP}} \circ f_{\mathrm{LogEig}} \circ f_{\mathrm{ReEig}} \circ f_{\mathrm{BiMap}} \circ f_{\COV} \circ f_{\mathrm{delay}}(\x),
\end{equation}
where each \( f_{l} \) represents a specific transformation or layer within the network. The learning process is formalized as the mapping \( f_{\methodmath}: \mathcal{X} \rightarrow \mathcal{Y} \), which operates on the training dataset. \( \theta \) denotes the set of parameters within the parameter space \( \Theta \), specifically the one of the BiMap and the MLP layer. The optimization objective is to minimize an average loss \( \ell \) over the training dataset, defined as:
\begin{equation}\label{eq:learning}
    \min_{\theta} \frac{1}{N} \sum_{i=1}^{N} \ell(f_{\methodmath}(\x_i), \mathbf{y}_i).
\end{equation}
The model's generalization capability is further assessed using an independent test set.

In a DL typical training process, optimization methods like Adam~\cite{kingma2014adam} are normally used for implementing the back-propagation procedure. However, in the context of SPDNet, we must learn the weights of $W$ of the BiMap layer in such a way that the new weights still belong to the Stiefel manifolds, i.e., they are still orthonormal matrices. To solve the problem, we use the RiemannAdam~\cite{becigneul2018riemannian} optimization from the geoopt~\cite{kochurov2020geoopt} library.
In addition to the optimization strategy, we use the standard cross-entropy loss function \( \ell \) as a loss function, a well-established and widely used objective function in classification tasks.

To summarize, the resulting architecture \method uses enriched SPD matrices that encapsulate a broader spectrum of information than their traditional counterparts thanks to the additional layer that implements the phase space reconstruction. Such an approach allows, through the use of nonlinear systems theory, the reconstruction of a phase space that contains more information with respect to the one extracted from the original signal, allowing the use of fewer electrodes. The augmented SPD matrices require an SPDNet with a larger number of parameters, which is partly counteracted by using fewer electrodes. In particular, this representation accentuates how the input SPD matrix is modified from the initial dimension $\mathcal{S}_{C}^{++}$ to $\mathcal{S}_{C \times \psi}^{++} $. 

\begin{figure}[!ht]
    \centering
    \includegraphics[width=\linewidth]{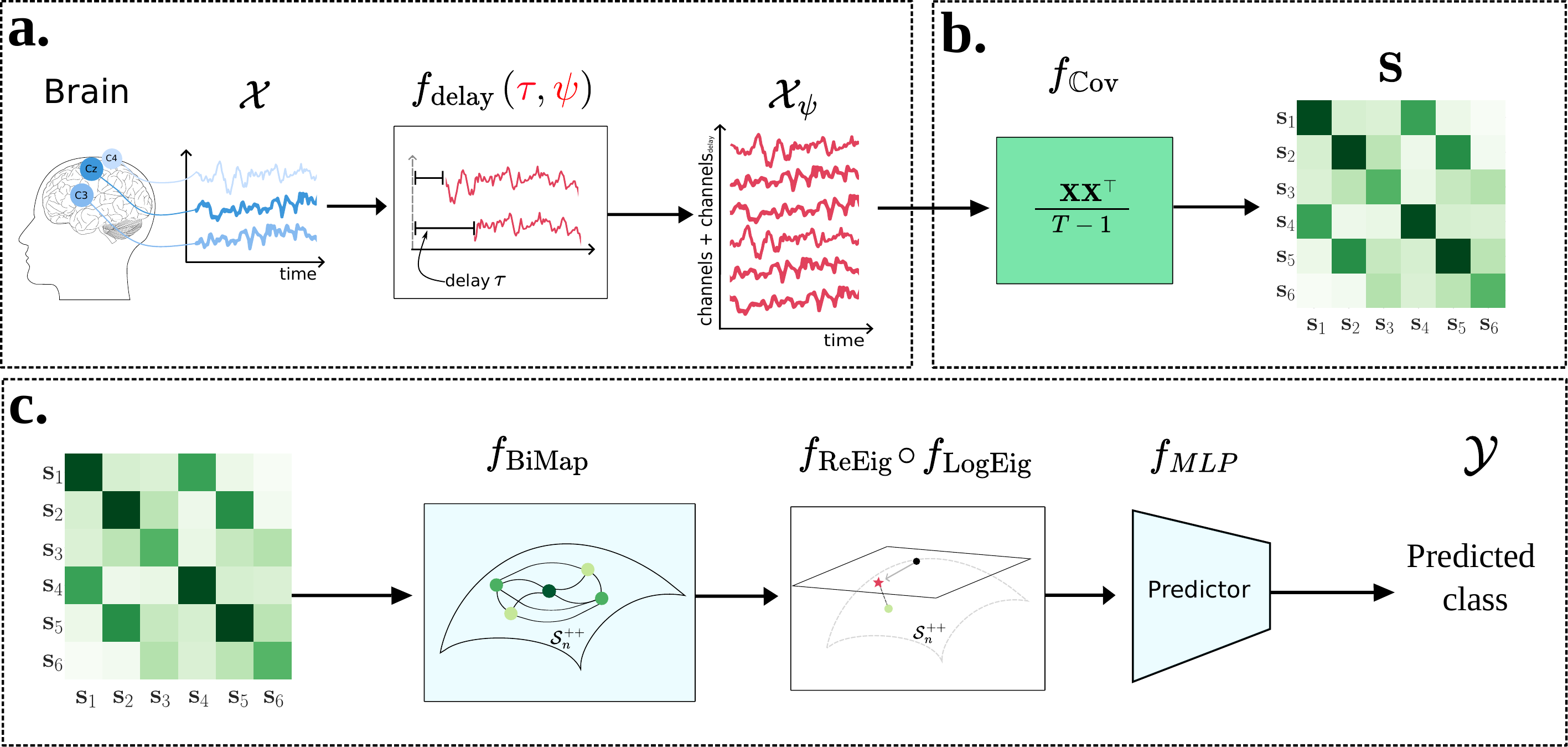}
    \caption{\textit{Overview of our approach.} \textbf{a.}~Phase space reconstruction for the EEG signal. For each trial from $\x$, we apply the function $f_{delay}$ to reconstruct the phase space. During each evaluation fold, we adopted the MDOP algorithm or optimized search to estimate suitable function parameters for embedding dimension $\psi$ and delay $\tau$. These hyperparameters are marked in red in the figure. For illustrative purposes, this figure considers 3 electrodes located over the motor cortex with parameters $\psi=2$ and $\tau=10$. \textbf{b.}~Conversion of the phase space time-series with $f_{\COV}$ for the covariance representation of the SPD, represented in green. With the parameter considered, the covariance is a square matrix of dimension 6. \textbf{c.}~The covariance feature space is then used to train the Symmetric Positive Definite Neural Network with BiMap, ReEig, LogEig, and MLP layers. Once the representation returns to the Euclidean space (red), after the LogEig, we adjust a fully connected layer to predict the associate label $\y$ for each trial. The light blue color represents the layer that possesses trainable parameters.}
    \label{fig:overview}
\end{figure}

Overall, this enrichment of the information contained in the SPD matrices not only expands the range of discriminative information that can be extracted but also forces the network to adapt to more intricate and fuzzy patterns in the data. \autoref{fig:overview} provides a graphical picture of our methodology.

In the case of \method, we set the subspace dimension of the BiMap layer at half of the input dimension~\cite{huang2017riemannian}. For standard SPDNet, we maintained the subspace dimension at the same value as the original input dimension (although in this case, the BiMap layer acts only as a rotation) in order to proceed with subsequent analyses concerning the explainability of the model. Anyway, diminishing the subspace dimension of half in the case of standard SPDNet consistently led to a reduction in performance. Note that the current implementation of SPDNet has a ReEig layer that is not scale-independent because of the absolute parameter $\varepsilon=10^{-4}$. To enhance the architecture's scale independence, we have introduced a standardization procedure for the raw signal, bringing every channel to a zero mean and a unit standard deviation. 

In short, Phase SPDNet processes the SPD matrix that contains the spatial information and the temporal evolution of the signal, while SPDNet handles only the spatial covariance.

\subsection{Datasets}
In order to assess the replicability of our approach, we used six different open datasets from \textsc{MOABB}~\cite{aristimunha-carrara-etal:23} consisting of almost 100 subjects as shown in Table~\ref{table:dataset}.

\begin{table}[!ht]
\resizebox{\linewidth}{!}{
\begin{tabular}{l|c|c|c|c|c|c|c}
\toprule
\textbf{ Dataset } & \textbf{ Subjects } & \textbf{ Channels } & \textbf{Sampling Rate (Hz)} & \textbf{ Sessions } & \textbf{ Tasks } & \textbf{ Trials/Class } & \textbf{Epoch (s)} \\
  \midrule
 \text{BNCI2014001~\cite{tangermann2012review}} & 9 & 22 & 250 & 2 & 4 & 144 & [2, 6] \\
 \text{BNCI2014004~\cite{leeb2007brain}} & 9 & 3 & 250 & 5 & 2 & 360 & [3, 7.5]\\
 \text{Cho2017~\cite{cho2017eeg}} & 52 & 64 & 512 & 1 & 2 & 100 & [0, 3]\\
 \text{Schirrmeister2017~\cite{schirrmeister2017deep}} & 14 & 128 & 500 & 1 & 4 & 120 & [0, 4]\\
 \text{Weibo2014~\cite{yi2014evaluation}} & 10 & 60 & 200 & 1 & 7 & 80 & [3, 7]\\
 \text{Zhou2016~\cite{zhou2016fully}} & 4 & 14 & 250 & 3 & 3 & 160 & [0, 5] \\
\bottomrule 
\end{tabular}
}
\caption{Motor Imagery datasets considered during this study}
\label{table:dataset}
\end{table}
We employed standard pre-processing steps designed for all datasets described in~\citep{aristimunha-carrara-etal:23}. The pre-processing steps are the same for all the pipelines considered in our study. We applied band-pass filtering between $[8-32]$\ Hz~\cite{nam2018brain, PFURTSCHELLER199765} and electrode standardization, bringing every channel to a zero mean and a unit standard deviation. We opted to use the complete epoch duration, although this duration varies among datasets, as shown in Table~\ref{table:dataset}.

We concentrated on a binary classification task, aiming to discern between imagined movements of the right and left hand using an intra-subject/within-session evaluation, which is based on a $5-$fold cross-validation conducted for each session. 

We also focus on investigating our algorithm's robustness with fewer electrode settings. For this purpose, our channel selection process was guided by two constraints: ensuring neurophysiological relevance and maintaining consistent electrode selection across datasets. We selected electrodes positioned over the sensorimotor region, particularly in the central (C) and centro-parietal (CP) lines, which are actively involved in the tasks. This choice is strongly supported by BCI research \cite{PFURTSCHELLER199765, MotorImagery2019, MUNZERT2009306, porro1996primary, NEUPER2006211, scherer2007sensorimotor, jeannerod1995mental}. In the considered datasets we used EEG montages with varying numbers of electrodes, from 3 to 128. We standardized our analysis by using the minimal overlapping electrodes common to all dataset: C3, Cz, and C4. All datasets considered were based on reference electrodes, and no average reference was used.

\subsection{Baseline Comparison}
\label{sec:Baseline}
We compare the performance of our pipeline to that of several state-of-the-art DL Neural Networks used for BCI. All algorithms in our study were tested using the same set of three electrodes to maintain uniformity in our experimental approach. To standardize the models' parameters, we initially resample the input time series to align the signal with the state of the art in order to use the parameter's architecture of the original paper.

\begin{enumerate}
    \item \textbf{ShallowNet}~\cite{schirrmeister2017},  a neural network architecture with independent spatial and temporal convolution steps with Relu activation, with standardized and re-sampled EEG signal at 250Hz.
    \item \textbf{DeepNet}~\cite{schirrmeister2017}, an approach similar to ShallowNet with a deeper linear layer, with standardized and re-sampled EEG signal at 250Hz.
    \item \textbf{EEGNet}~\cite{lawhern2018eegnet}, have a depth-wise convolutional layer functions as a spatial filter across channels, complemented by a separable convolution layer for features extraction designed for EEG with a sample frequency of 128Hz.
    \item \textbf{EEGTCNet}~\cite{ingolfsson2020eegtcnet}, is a neural network that wrapper the EEGNet and includes a Temporal Convolution Network over the embedded representation, with standardized and re-sampled EEG signal at 250Hz. 
    \item \textbf{EEGITNet}~\cite{salami2022eeg}, a neural network inspired by InceptionNet with parallel convolution layers with different scales. The network has been designed for EEG signals sampled at 128Hz.
    \item \textbf{EEGNeX}~\cite{chen2022toward}, a neural network inspired by EEGNet incorporation the key components from ConvNeXt. The network has been designed for EEG signals sampled at 128Hz. 
    \item \textbf{DynSpat-EEGNet}~\cite{banville2022robust}, a neural network based on Dynamic spatial filtering with the EEGNet as head. The network has been designed for EEG signals sampled at 250Hz.
    \item \textbf{DynSpat-ShallowNet}~\cite{banville2022robust}, a neural network based on Dynamic spatial filtering with the EEGNet as head. The network has been designed for EEG signals sampled at 250Hz.
\end{enumerate}

These methods were implemented and trained using \textsc{MOABB} version 1.1.0~\cite{aristimunha-carrara-etal:23}. 
For in-depth information about the deep learning hyper-parameters, please consult Table~\ref{table:pipeline_parameter_DL}.

\section{Results}
\label{sec:Results}
In this section, we analyze the results produced using within-session (WS) evaluation. This method is based on a 5-fold cross-validation performed on each session independently. In particular, we report two separate analyses. 

The first analysis is dedicated to contrasting the performance of \method, showing both MDOP and optimized search, with a specific focus on the exploitation of the covariance feature, against the state-of-the-art techniques in DL applied in EEG. The subsequent analysis, reported in the Appendix, shifts the focus towards the exploration of coherence as a feature for classification. Despite the ongoing investigations into coherence, we deliberately separated this analysis due to its comparatively lesser performance against the benchmarks achieved with covariance features. Nevertheless, the augmentation procedure is a significative methodology to improve the performance of both feature sets.

\begin{figure}[!ht]  
    \centering
    \centering
     \subfloat[]{%
            \includegraphics[width=0.6\linewidth]{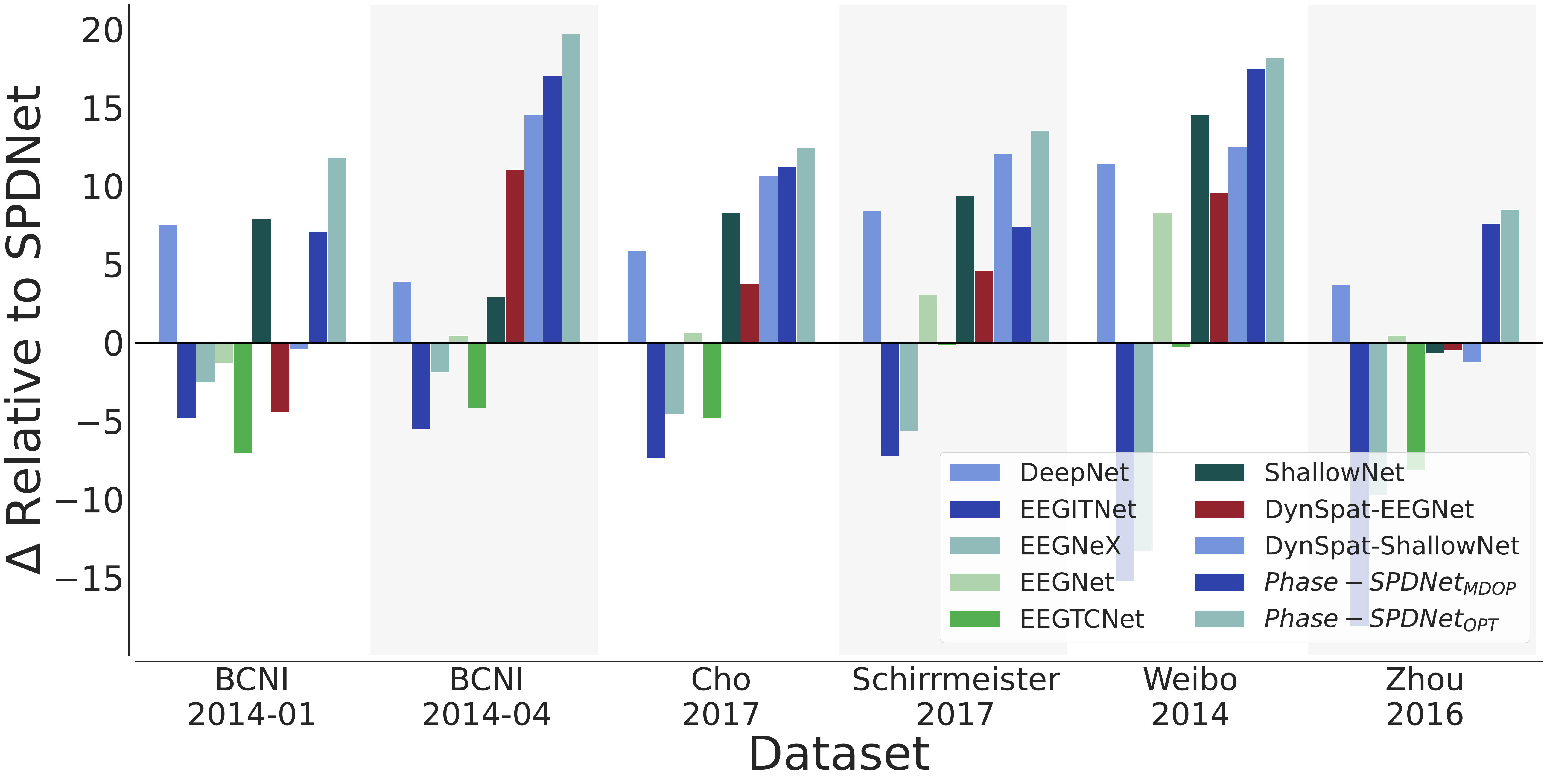}}
            \hfill
     \subfloat[]{%
            \includegraphics[width=0.37\linewidth]{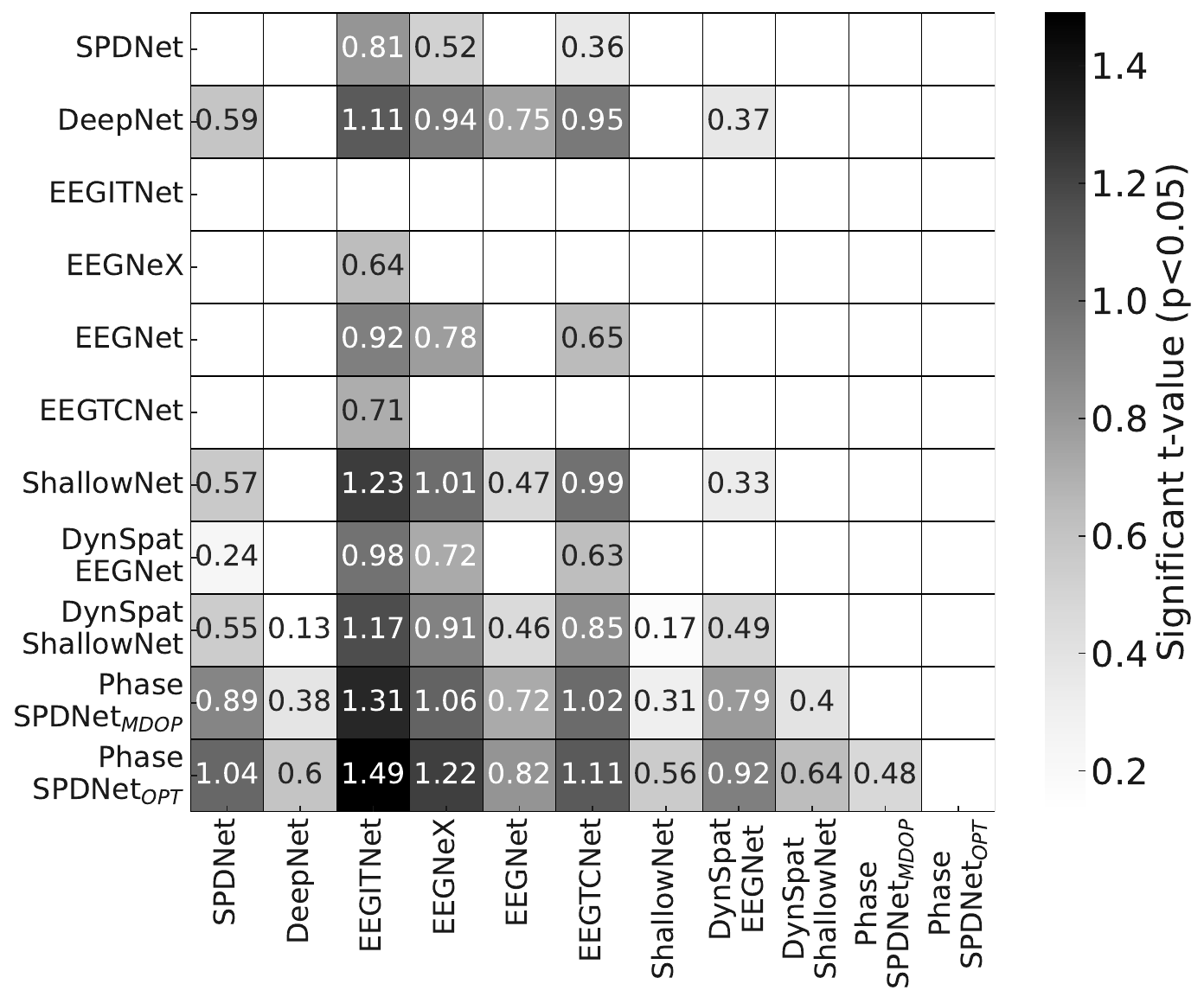}}
    \\
   \subfloat[]{%
            \includegraphics[width=0.33\linewidth]{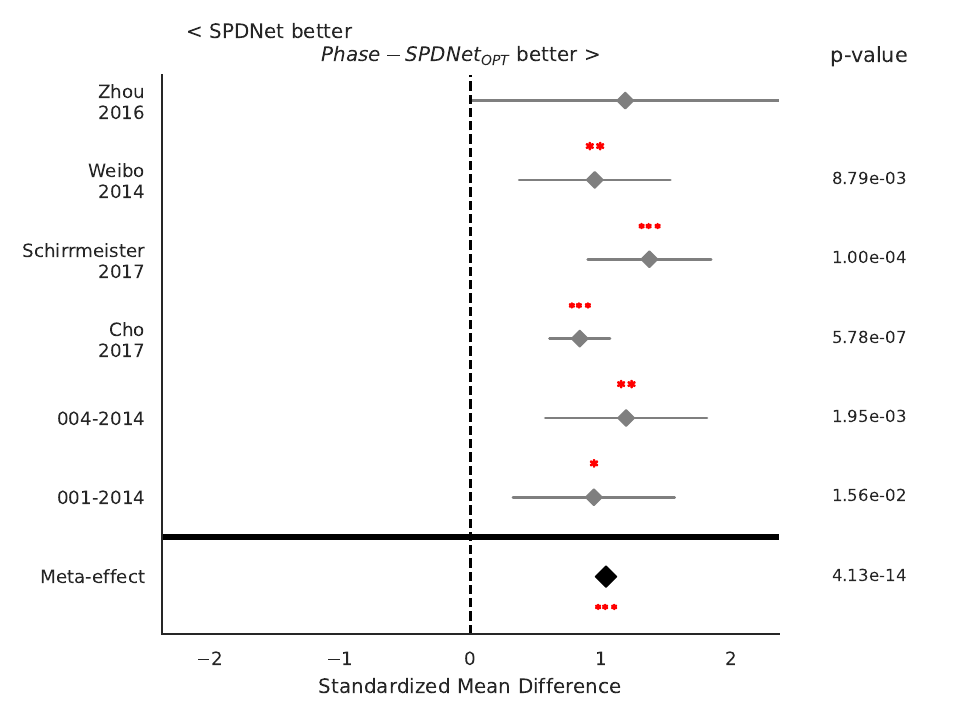}}
            \hfill
   \subfloat[]{%
            \includegraphics[width=0.33\linewidth]{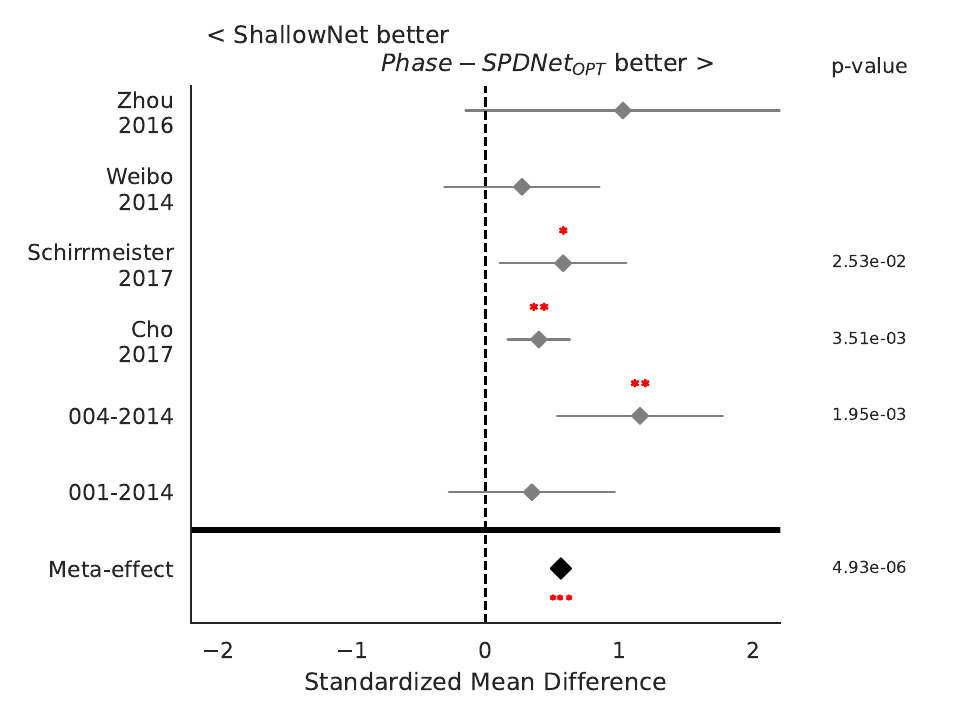}}
            \hfill
   \subfloat[]{%
            \includegraphics[width=0.33\linewidth]{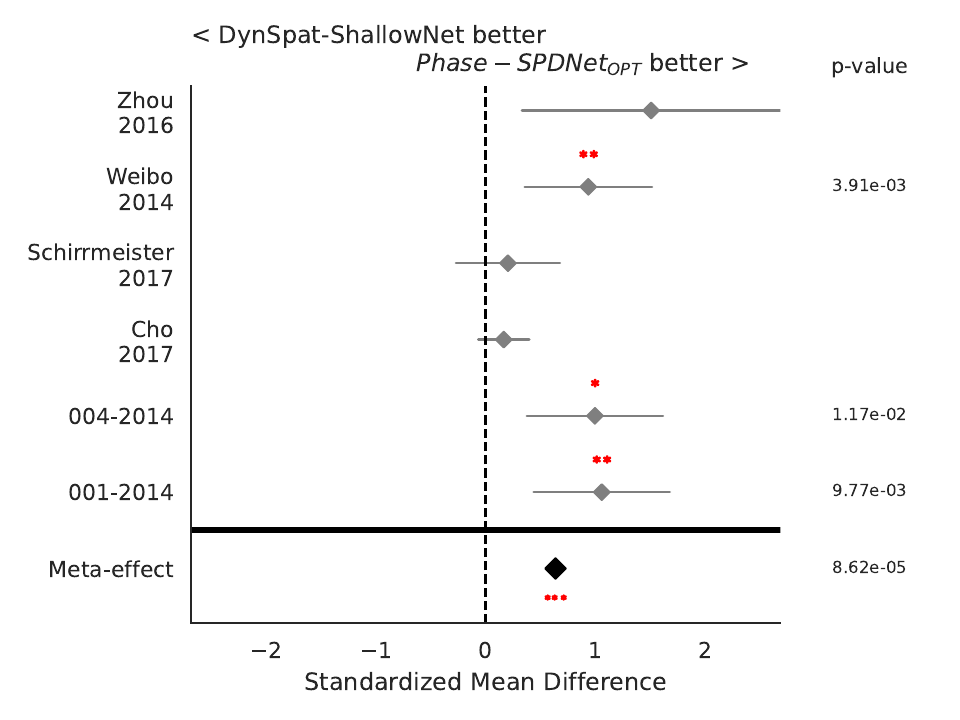}}
            \hfill

    \caption{Results for Right vs Left-hand classification, using Within-Session evaluation. Plot (a) provides the relative improvement of the AUC-ROC in percentage of the method considered with respect to the standard SPDNet of the different pipelines considered. Plot (b) shows a combined meta-analysis (over all datasets) of the different pipelines. It shows the significance of the algorithm on the y-axis being better than the one on the x-axis. The gray level represents the significance level of the ROC-AUC difference in terms of t-values.  We only show significant interactions ($p < 0.05$). Plots (c), (d), and (e) show the meta-analysis of \method$_{OPT}$ against SPDNet, ShallowNet, and DynSpat+ShallowNet, respectively. We show the standardized mean differences of p-values computed as a one-tailed Wilcoxon signed-rank test for the hypothesis given in the plot title. The {\color{gray}\textbf{gray}} bar denotes the $95\%$ interval. {\color{red}{\textbf{*}}} stands for $p < 0.05$, {\color{red}{\textbf{**}}} for $p < 0.01$, and {\color{red}{\textbf{***}}} for $p < 0.001$.
    }
    \label{results_fig}
\end{figure}

For the first analysis, we tested the performance of \method$_{MDOP}$ and \method$_{OPT}$ against the baseline comparison defined in \ref{sec:Baseline} across a broad spectrum of datasets. The results are listed in Table~\ref{table:rhlh_DL}, while the detailed statistical analysis can be found in Figure~\ref{results_fig}.

Table~\ref{table:rhlh_DL} shows ROC AUC results for right-hand versus left-hand classification, where bold numbers represent the best score in each dataset, revealing a consistent trend wherein our algorithm \method$_{OPT}$ surpasses the DL state-of-the-art in the context of reduced datasets in all considered datasets. The statistical analysis conducted, shown in Figure~\ref{results_fig} (b), allows us to demonstrate that \method$_{OPT}$ outperforms the state-of-the-art DL methodology in a reduced dataset context.

When employing MDOP, \method$_{MDOP}$ obtains suboptimal performance with respect to \method$_{OPT}$, finding aligned with current research~\cite{carrara2024classification}, while still surpassing the performance of the DL state-of-the-art. A notable exception is the Schirrmeister2017 dataset, where \method$_{MDOP}$ is marginally surpassed by ShallowNet, DeepNet, and DynSpat-ShallowNet. However, we are interested in computing a statistical behavior considering all datasets under consideration. When considering the aggregated results across all datasets, our meta-analysis shows that \method$_{MDOP}$ yields statistically better results than all other approaches, as shown in Fig.~\ref{results_fig} (b).

The second relevant point is to see the impact of the augmentation procedure compared to the standard SPDNet. Figure~\ref{results_fig} (a) elucidates this aspect by illustrating the relative performance enhancements of various models against the SPDNet standard. Notably, ShallowNet and DeepNet exhibit positive performance increments across the majority of the datasets examined. However, the augmentation procedure, as implemented by \method$_{OPT}$, not only consistently yields positive relative improvement against the standard SPDNet across all datasets but also the improvement introduced by the augmentation procedure turns out to be so significant that it even outperforms methods that initially surpassed the SPDNet benchmark. This improvement is not only demonstrated by the relative improvement with respect to the SPDNet standard but also is evident through the meta-analysis comparison (refer to Fig~\ref{results_fig} (c), (d) and (e)).

Ultimately, the versatility and reliability of \method$_{OPT}$, particularly in data-limited scenarios, position the method as a promising solution for real-world applications.

\begin{table}[!ht]
\centering
\caption{Average AUC-ROC (\%) on six datasets (BCNI2014001, BCNI2014004, Cho2017, Schirrmeister2017, Weibo2014, and Zhou2016), for a left vs. right motor imagery task. The highest performances are in bold-face. The gray area marks the second to best pipeline excluding the \method$_{MDOP}$.}
\resizebox{1.0\textwidth}{!}{
\begin{tabular}{l|c|c|c|c|c|c} 
\toprule 
  \multirow{1}{*}{\bf Models} & \multicolumn{1}{|c|}{\bf BNCI2014001} & \textbf{BNCI2014004} & \textbf{Cho2017} & \textbf{Schirrmeister2017} &               \textbf{Weibo2014} & \textbf{Zhou2016}\\
  \midrule
DeepNet~\cite{schirrmeister2017} (2017) & 75.80 $\pm$ 15.45 & 72.80 $\pm$ 19.48 & 63.13 $\pm$ 14.45 & 73.04 $\pm$ 15.67 & 73.97 $\pm$ 18.07 &  \cellcolor{gray!25}91.74 $\pm$ 7.00 \\
ShallowNet~\cite{schirrmeister2017} (2017) & \cellcolor{gray!25}75.85 $\pm$ 15.41 & 72.17 $\pm$ 18.61 & 64.14 $\pm$ 13.03 & 73.59 $\pm$ 15.19 &  \cellcolor{gray!25}75.36 $\pm$ 15.69 & 88.03 $\pm$ 8.55 \\
EEGNet~\cite{lawhern2018eegnet} (2018)& 70.64 $\pm$ 19.87 & 70.27 $\pm$ 18.91 & 60.23 $\pm$ 14.98 & 69.80 $\pm$ 16.51 & 71.94 $\pm$ 17.80 & 88.95 $\pm$ 7.84 \\
EEGTCNet~\cite{ingolfsson2020eegtcnet} (2020) & 65.98 $\pm$ 17.25 & 66.86 $\pm$ 18.41 & 56.43 $\pm$ 12.31 & 67.87 $\pm$ 17.36 & 65.94 $\pm$ 15.69 & 81.47 $\pm$ 11.66 \\
EEGITNet~\cite{salami2022eeg} (2022) & 66.64 $\pm$ 13.98 & 64.93 $\pm$ 14.49 & 54.68 $\pm$ 11.97 & 62.98 $\pm$ 16.19 & 56.97 $\pm$ 17.66 & 72.82 $\pm$ 12.78 \\
EEGNeX~\cite{chen2022toward} (2022) & 68.86 $\pm$ 17.27 & 68.29 $\pm$ 17.85 & 56.64 $\pm$ 12.83 & 64.26 $\pm$ 17.58 & 58.73 $\pm$ 19.40 & 80.25 $\pm$ 15.55 \\
DynSpat+EEGNet~\cite{banville2022robust} (2022) & 68.06 $\pm$ 18.58 & 76.61 $\pm$ 16.81 & 61.51 $\pm$ 13.74 & 70.62 $\pm$ 16.53 & 72.88 $\pm$ 19.33 & 87.85 $\pm$ 7.01 \\
DynSpat+ShallowNet~\cite{banville2022robust} (2022) & 70.95 $\pm$ 18.78 &  \cellcolor{gray!25}78.95 $\pm$ 17.23 &  \cellcolor{gray!25}65.52 $\pm$ 14.65 &  \cellcolor{gray!25}75.07 $\pm$ 14.40 & 74.70 $\pm$ 19.12 & 87.33 $\pm$ 8.87 \\
SPDNet~\cite{huang2017riemannian} (2017) & 71.02 $\pm$ 15.64 & 70.15 $\pm$ 16.85 & 59.95 $\pm$ 12.61 & 67.40 $\pm$ 13.01 & 67.04 $\pm$ 17.62 & 88.85 $\pm$ 8.05 \\
\method$_{MDOP}$ (Our) & 75.98 $\pm$ 17.00 & 80.46 $\pm$ 16.64 & 66.00 $\pm$ 13.29 & 72.02 $\pm$ 13.07 & 78.01 $\pm$ 19.64 & 94.92 $\pm$ 3.34 \\
\method$_{OPT}$ (Our) & \textbf{78.40 $\pm$ 15.12} & \textbf{82.29 $\pm$ 17.04} & \textbf{66.69 $\pm$ 12.99} & \textbf{76.05 $\pm$ 13.84} & \textbf{78.18 $\pm$ 19.36} & \textbf{95.62 $\pm$ 2.36} \\
\bottomrule
\end{tabular}}
\label{table:rhlh_DL}
\end{table}

\section{Discussion}
\label{sec:Discussion}

\subsection{Impact of augmentation on most and least responsive 5 subject of Cho2017}
In the previous section, we explored the outcome obtained by \method{} within the context of a reduced number of electrodes. Now, our attention shifts to a deeper exploration of this new algorithm. Still, in the context of three electrodes, we have chosen to concentrate on the examination of the five most and least responsive subjects within the Cho2017 dataset (the wider dataset in terms of the number of subjects). These subjects were selected on the basis of the MDM algorithm~\cite{barachant2010riemannian} with covariance as a feature.

\begin{figure}[!ht]
\centering
       \includegraphics[width=1\linewidth]{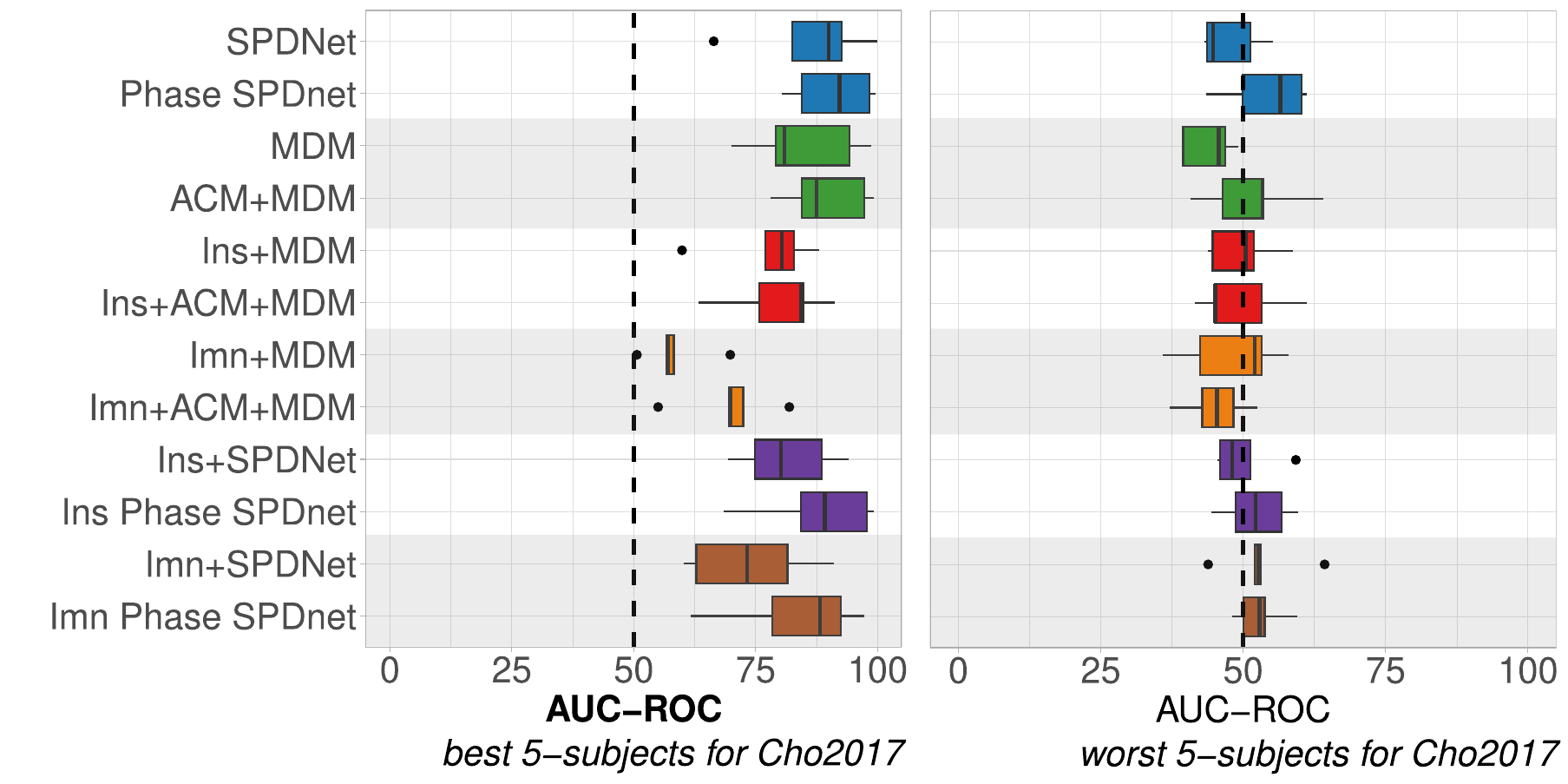}

\caption{Most and least responsive 5 subjects of Cho2017. (a) Plot showing the box plot of the five most responsive subjects (3, 14, 35, 41, 43) of Cho2017. We see that the augmentation procedure consistently leads to notable improvements in the classification performance for all the SPD estimators considered. (b) Plot showing the box plot of the least responsive five subjects (2, 7, 29, 34, 50) of Cho2017. We see that the augmentation procedure increases the performance for the covariance feature, while the improvement is less clear for some FC estimators. In the plot, the names Ins and Imn stand respectively for Instantaneous Coherence and Imaginary Coherence.
}
\label{fig:BoxPlotBestWorst}
\end{figure}

To analyze the performance, we use different pipelines, MDM~\cite{barachant2010riemannian}, ACM+MDM~\cite{carrara2024classification}, SPDNet, and \method$_{OPT}$, each offering a distinct perspective on the impact of augmentation techniques on both traditional ML and advanced DL algorithms. Furthermore, we assess the performance using three different SPD features: covariance, instantaneous coherence, and imaginary coherence (Figure~\ref{fig:BoxPlotBestWorst}).

For the five most responsive subjects, covariance stands out as the top-performing feature, as could be expected. However, in the case of the five least responsive subjects, this dominance of covariance is less clear when considering the state-of-the-art methodology. Nevertheless, considering the \method$_{OPT}$ method that uses covariance as a feature, we see that covariance is again the best-performing feature.

The outcomes of our study are in concordance with the state-of-the-art~\cite{fucone:2022}: results achieved using Covariance as SPD feature outperform those obtained with the Functional Connectivity (FC) estimator.

The augmentation procedure consistently leads to notable improvements in the classification performance for the top-performing five subjects.  However, the situation becomes less definitive for the least responsive five subjects, where the augmentation effect is evident for the covariance but is not as pronounced for the FC estimator. In this case, their signals exhibit a lower SNR, which can substantially impact the MDOP algorithm, leading to a poor estimation of the hyper-parameter.

\subsection{Interpretability}
For analyzing the explainability of \method{}, focusing on the best performing \method$_{OPT}$, we use the GradCam++ algorithm~\cite{chattopadhay2018grad}, a technique designed for visualizing and explaining the decision-making process of neural networks. The GradCam++ uses the derivative of the gradient for a specific target (e.g., right hand or left hand) that flows up to a specific convolutional layer. This process creates a map highlighting the critical regions used to predict that specific target class, without requiring architectural modifications or re-training. GradCam++ has proven to be a robust method for spatial interpretability of EEG signals in a class-agnostic manner~\cite{sujatha2023empirical}.

\begin{figure}
\centering
    \includegraphics[width=\linewidth]{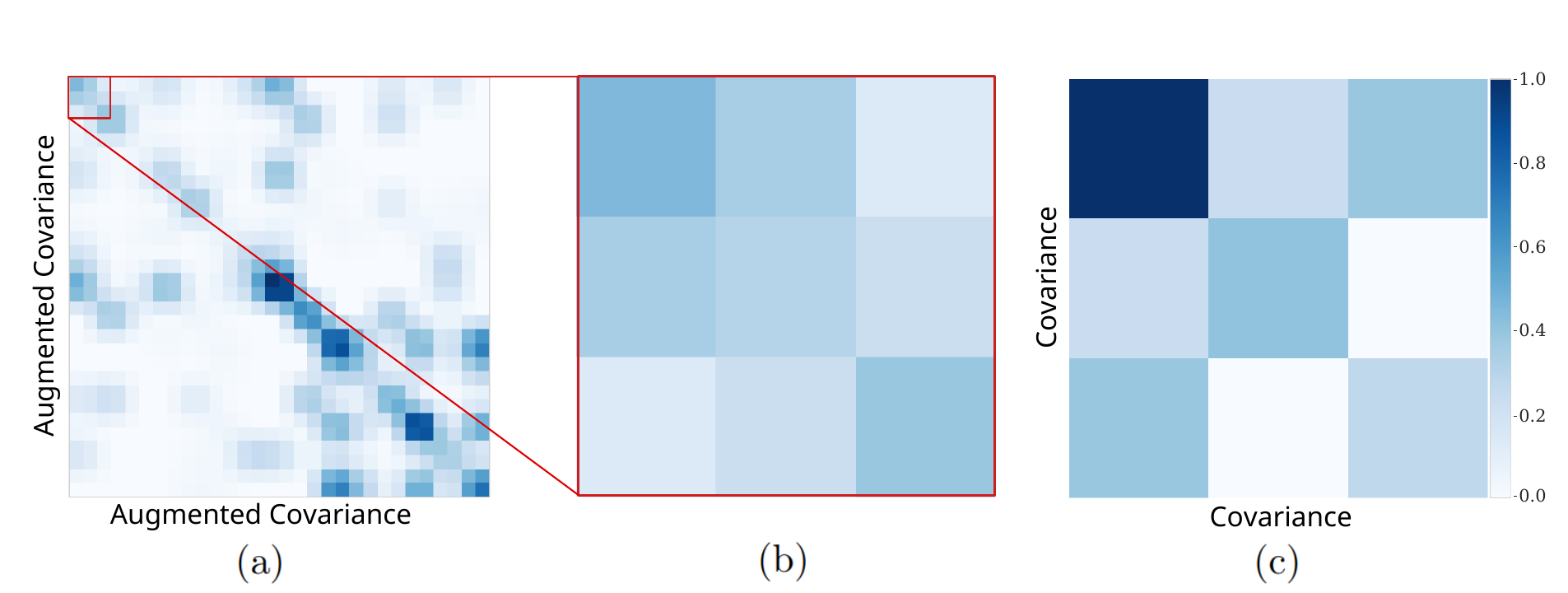}
    \caption{Interpretability parallel between GradCam++ and t-test for right versus left-hand classification for subject 35 of Cho2017. 
    Plot (a) is the GradCam++ saliency map obtained at the level of the ReEig layer, for \method$_{OPT}$ with a target left hand. Plot (b) is a zoom in the region corresponding to standard Covariance. Plot (c) the GradCam++ saliency map obtained at the level of the ReEig layer, for SPDNet with a target left hand. This discrepancy highlights the added value of our method in identifying crucial inter-channel relationships that might be overlooked by traditional covariance analysis.}
    \label{fig:GradCam_CovSPDNet}
\end{figure}

In our analysis, we focus on Subject 35 from the Cho2017 dataset, which is notable for being among the top 5 subjects that exhibit the most substantial performance improvement through the augmentation procedure. Similar findings also hold for other subjects. We create the GradCam++ map, produced with left hand as a target, on a sample of the test dataset that was correctly classified by the \method$_{OPT}$ algorithm (left hand classified as left hand) but was misclassified by SPDNet using the same feature, as shown in Figure~\ref{fig:GradCam_CovSPDNet}. The obtained map represents the GradCam++ obtained at the ReEig layer.

Standard SPDNet places more emphasis on the diagonal terms of SPD matrices. On the other side, when employing \method$_{OPT}$, a more comprehensive picture unfolds. While diagonal terms continue to play a pivotal role, what becomes increasingly evident is the significant contribution of some off-diagonal terms. Moreover, the \method with optuna emphasizes the interplay between a channel and its lagged version, showing that this information is even more relevant for the classification with respect to the standard covariance itself.

In essence, this observation emphasizes that the \method$_{OPT}$ model leverages both individual electrode characteristics (diagonal terms) and the interplay between electrodes in time and space (off-diagonal terms) as crucial elements in its classification decision-making process.


\subsection{Convergence Behavior}
\begin{figure}[ht]  
    \centering
    \centering
     \subfloat[]{%
            \includegraphics[width=0.5\linewidth]{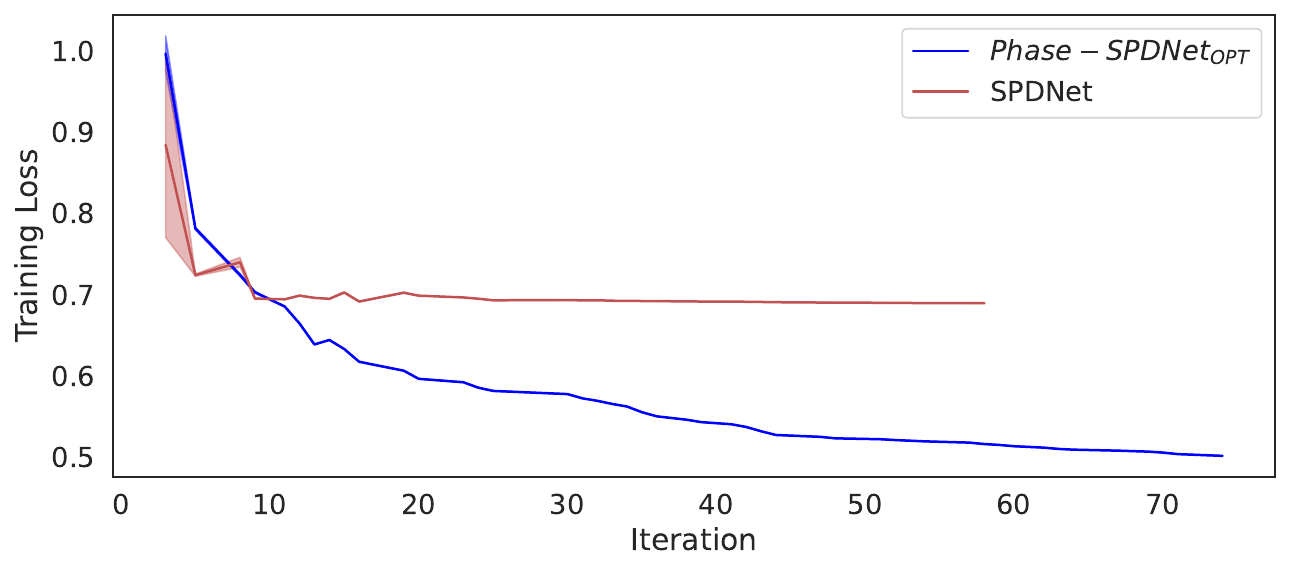}}
            \hfill
     \subfloat[]{%
            \includegraphics[width=0.5\linewidth]{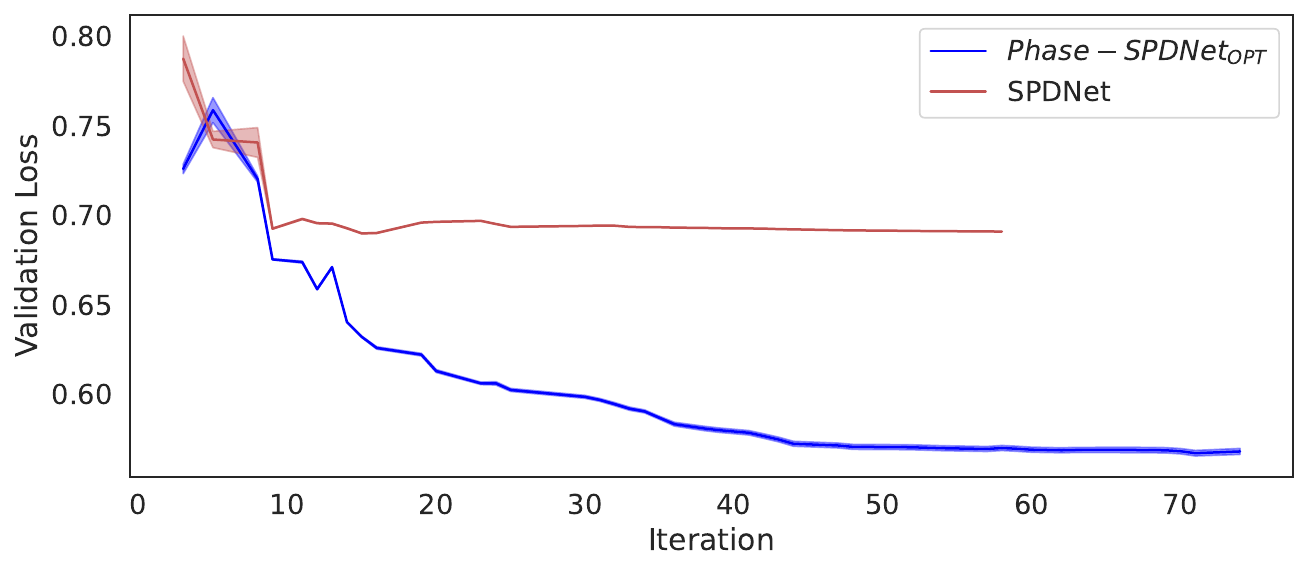}}
            \hfill
    \caption{Convergence behavior for right-hand vs. left-hand classification using WS evaluation. The plot uses subject 35 of the Cho2017 dataset, showing the mean value for each plot computed over the 5-Fold considered in the WS evaluation. Plot (a) provides the training loss, while plot (b) is the validation one.
    }
    \label{Convergence}
\end{figure}

To analyze convergence behavior, we again consider subject 35 of the Cho2017 dataset. We examine the outcomes concerning both training and validation loss. The consolidated results are visually presented in Figure~\ref{Convergence}, wherein the graphs depict the average values across the considered 5-fold cross-validation.

Notably, our observations reveal that the loss function of \method$_{OPT}$ shows a rapid decline in the loss function during the initial iteration, suggesting it learns more quickly than the SPDNet model. Moreover, \method$_{OPT}$ steadily decreases the loss, while the SPDNet decreases initially but plateaus around a higher value compared to the proposed approach. These findings highlight that \method$_{OPT}$ possesses a better learning efficiency and generalization capability.

\subsection{Trainable Parameters}
The proposed methodology has demonstrated a statistically significant enhancement in performance without compromising efficiency. Indeed, this model features a reduced number of trainable parameters, particularly influenced by the hyperparameter $\psi$, as shown in Table~\ref{tab:Architecture}. The table offers an in-depth analysis of how variations in $\psi$ affect the overall architecture.

While there is an observable increase in the number of trainable parameters when compared to the conventional SPDNet, this increase is balanced by a significant improvement in performance.  Despite the additional parameters relative to the SPDNet standard, our model remains considerably more efficient in terms of parametric complexity than the current state-of-the-art DL pipelines, as shown in Figure~\ref{fig:Param}. 

\begin{minipage}[b][][t]{\textwidth}
  \begin{minipage}[t]{0.49\textwidth}
    \centering
   \includegraphics[width=0.99\linewidth]{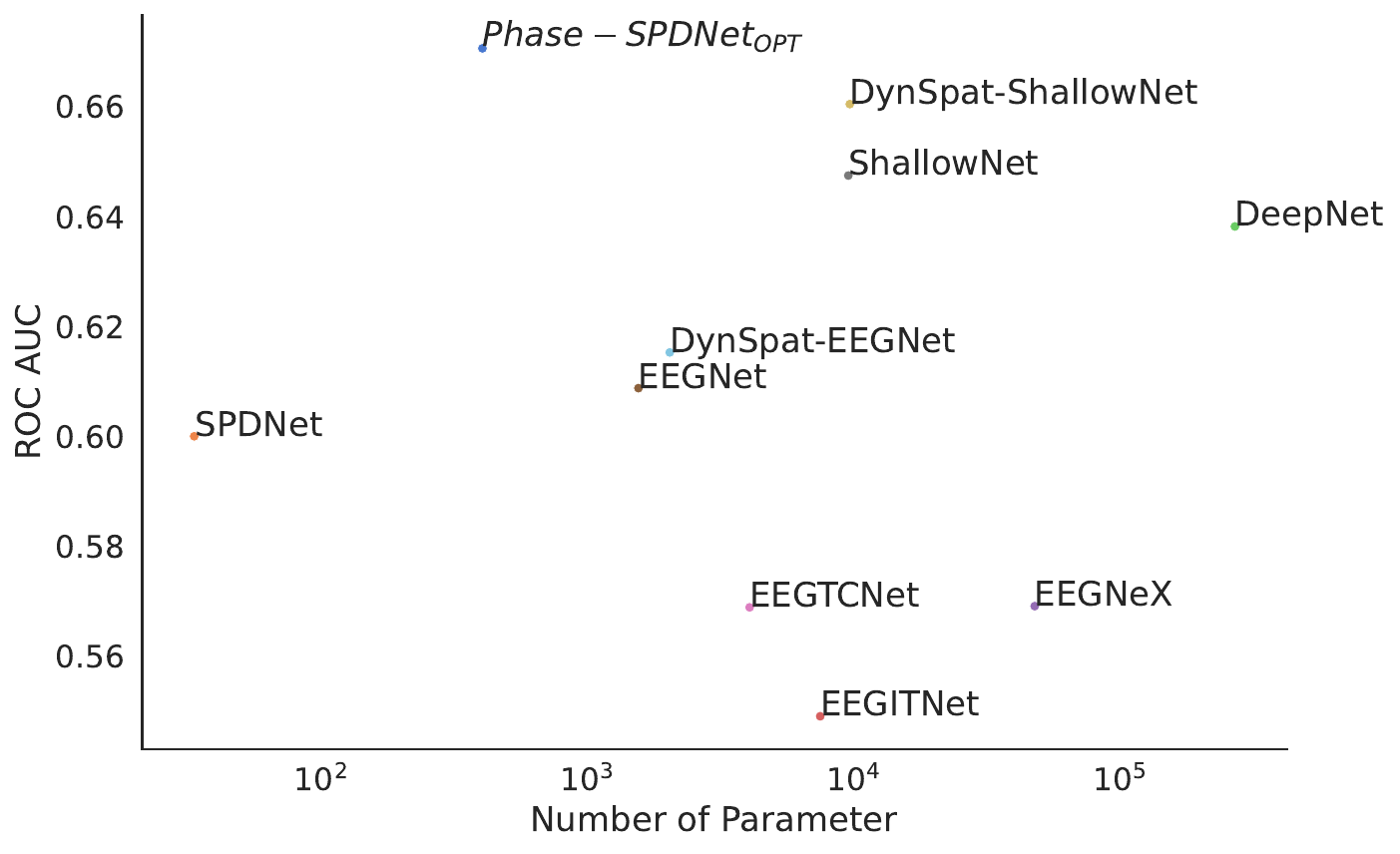}
    \captionsetup{type=figure}
    \captionof{figure}{Performance (ROC AUC) versus number of trainable parameters (logarithmic scale) for the Cho2017 dataset.}
    \label{fig:Param}
  \end{minipage}
  \hfill
  \begin{minipage}[b]{0.49\textwidth}
    \centering
    \resizebox{\linewidth}{!}{
    \begin{tabular}{l l r r}
         Stage & Layer & Output & Parameters \\
         \hline
         Input & & [1, 3, T] & 0 \\
         Augmentation & AugmentedDataset & [1, $3 \times \psi$, $T-\psi \tau$] & 0 \\
         Covariance & Covariances & [1, $\beta$, $\beta$] & 0 \\
         SPDNet & BiMap &  [1, $\alpha$, $\alpha$] & $\alpha * \beta$\\
         & ReEig &  [1, $\alpha$, $\alpha$] & 0 \\
         & LogEig &  [1, $(\alpha*(\alpha+1))/2$] & 0 \\
         Classification & Linear &  [1, 2] & $(\alpha*(\alpha+1))/2$ \\
         \hline
         Total & & & $(\alpha * \beta) + (\alpha*(\alpha+1))/2$
    \end{tabular}
    }
     \captionsetup{type=table}
    \captionof{table}{Architecture of \method with parameter number depending on $\psi$. \newline
    Using $\beta = 3 \times \psi$ and $\alpha = (3 \times \psi)/2$.}
    \label{tab:Architecture}
    \end{minipage}
    \label{tab:comp}
  \end{minipage}

\subsection{Limitation and Future Directions} 
Further enhancements could be achieved by adopting more intricate SPDNet architectures, such as incorporating batch normalization, exploring different BiMap layers, or experimenting with a Bottleneck architecture.

An intriguing direction arises from the insights gained in the explainability study, revealing that \method$_{OPT}$ predominantly focuses on diagonal elements. This suggests the possibility of employing a Region of Interest approach, where the network initially prioritizes the main diagonals while treating the remainder of the matrix as background.

Continued improvements could be realized through comprehensive algorithm testing across diverse datasets, tasks, and evaluation procedures. This approach allows for a better understanding of performance in more complex scenarios, particularly those posed by intra- and inter-subject variabilities.

Using three electrodes only is advantageous for practical scenarios. In fact, this decision was strategically made to shorten the length of the experiment, thereby minimizing patient fatigue and maintaining their attention throughout the session. 

Another challenge was the difficulty in classifying data from less responsive subjects. Despite our efforts, there remains a need for improved methods to better capture the subjects' intentions and their specific characteristics. 

\section{Conclusion}
\label{sec:Conclusion}
This study focuses on integrating the Augmented Covariance Matrix (ACM) with SPDNet to develop a practical algorithm intended for real-world applications, with a deliberate focus on scenarios involving a limited number of electrodes. Specifically, we restricted our electrode usage to just three, strategically placed on the motor cortex.

In fact, the ACM, based on Takens' theorem, demonstrates a particularly effective performance when applied with a reduced number of electrodes. However, it is important to note that using only three electrodes can impact the results. For instance, in the BNCI2014001 dataset, employing just three electrodes led to a performance reduction of about $10\%$ in the ShallowNet architecture compared to using all available electrodes. We consider ShallowNet as it was the best performing DL model in this task~\cite{chevallier2024largest}.

Validation of our approach involves leveraging nearly 100 subjects from several open datasets, a task facilitated by the MOABB framework. The resulting \method algorithm outperforms existing DL algorithms in BCI-EEG classification and offers some explainability elements through GradCam++ visualization. Remarkably, this algorithm requires a modest number of trainable parameters.

\section{Acknowledgments}
\label{sec:Acknowledgments}
The work of IG and TG was partly funded by the EUR DS4H/Neuromod fellowship. The BA work was supported partly by CAPES under Grant 001 and by DATAIA Convergence Institute as part of the ``Programme d’Investissement d’Avenir'', (ANR-17-CONV-0003) operated by LISN. We extend our gratitude towards the OPAL infrastructure at Université Côte d'Azur for their essential resources and support. Additionally, acknowledgment is due for the support from the European Research Council (ERC) under the EU’s Horizon 2020 research and innovation program (grant No. 864729), and the “Investissements d’avenir” program ANR-10-IAIHU-06.

\section{Open-source availability}
\label{sec:code}
The codes used to produce the results of this study are publicly available in this Github repository: 
\url{https://github.com/carraraig/Phase-SPDNet}. 
  
\bibliographystyle{unsrtnat}  
\bibliography{biblio}

\clearpage

\appendix
\setcounter{table}{0}
\renewcommand{\thetable}{A\arabic{table}}

\setcounter{figure}{0}
\renewcommand{\thefigure}{A\arabic{figure}}

\section{Results using Coherence as feature}
\label{sec:Appendices}

In this section, we report results using Imaginary and Instantaneous coherence features as input. Specifically, we aim to draw a comparison between the \method, showing both MDOP and optimized hyperparameter search, results and the non DL state-of-the-art pipelines (usually created for covariance but here applied on coherence). It's worth noting that, in the subsequent analysis, we abstain from comparing these results with the performance of state-of-the-art DL models, given that the employment of coherence-based features typically yields lower results across the board. Nevertheless, we find it particularly intriguing that both \method$_{OPT}$ and \method$_{MDOP}$ demonstrate a statistically significant improvement. This shows the potential and significance of \method in the broader landscape of signal processing and pattern recognition. We outline below the pipeline used for comparison with the coherence features-based approach:

\begin{enumerate}
    \item \textbf{CohCSP+LDA}~\cite{lotte2010learning}, a combination of the Common spatial pattern (CSP) algorithm followed by a classification performed on a shrinkage Linear Discriminant Analysis (LDA).
    \item \textbf{Coh + TS + EN}~\cite{fucone:2022}, a Riemannian classification method in the tangent space using an Elastic Network as classifier.
    \item \textbf{CohFgMDM}~\cite{yger2016riemannian}, a Riemannian classification method by Minimum Distance to the Mean after having applied a geodesic filtering.
    \item \textbf{CohMDM}~\cite{barachant2010riemannian}, a Riemannian classification method by Minimum Distance to the Mean.
    \item \textbf{Coh + TS + SVM}~\cite{barachant2010riemannian}, a Riemannian classification method in the tangent space using a Support Vector Machine (SVM).
\end{enumerate}

\begin{figure*}[ht]  
    \centering
    \centering
    \subfloat[]{%
                        \includegraphics[width=0.65\linewidth]{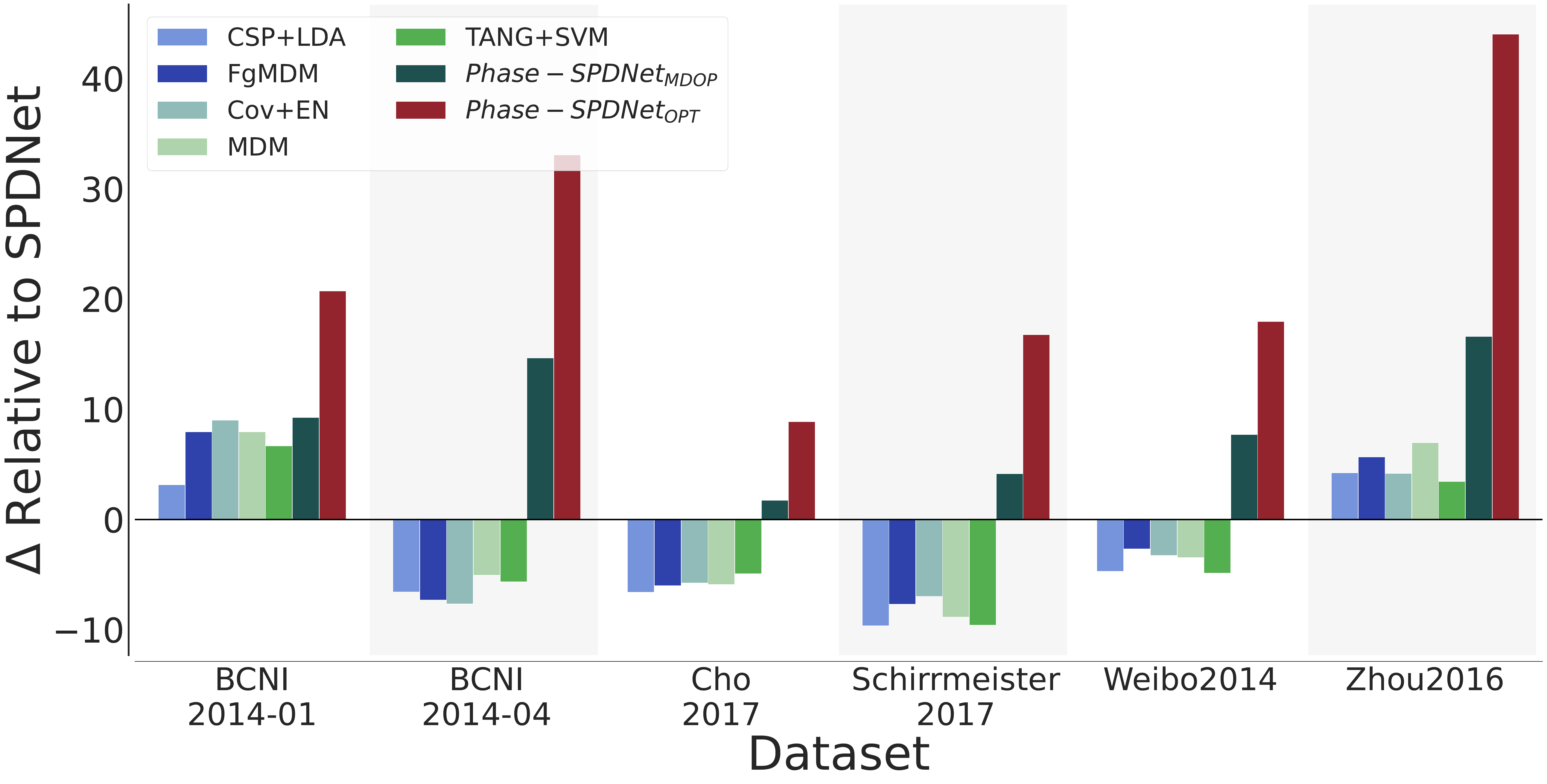}}   
                                    \hfill
    \subfloat[]{%
            \includegraphics[width=0.33\linewidth]{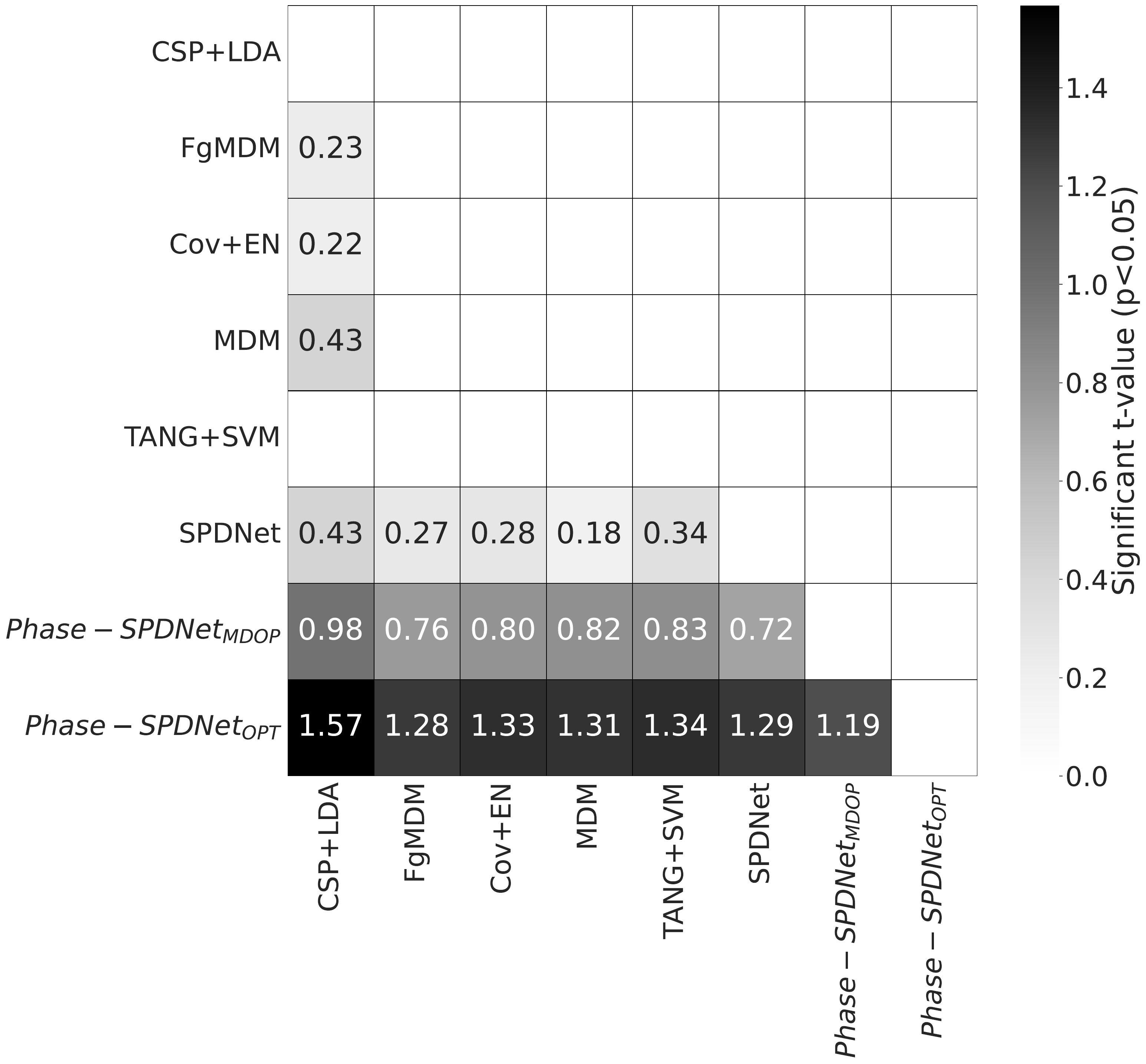}}  \\
   \subfloat[]{%
            \includegraphics[width=0.33\linewidth]{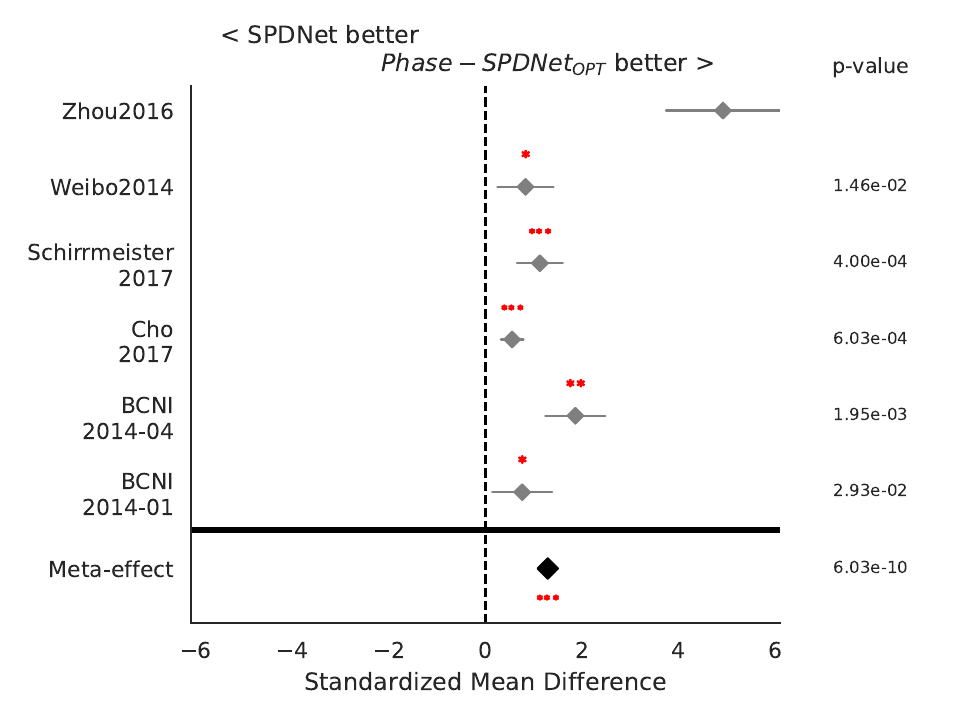}}
            \hfill
   \subfloat[]{%
            \includegraphics[width=0.33\linewidth]{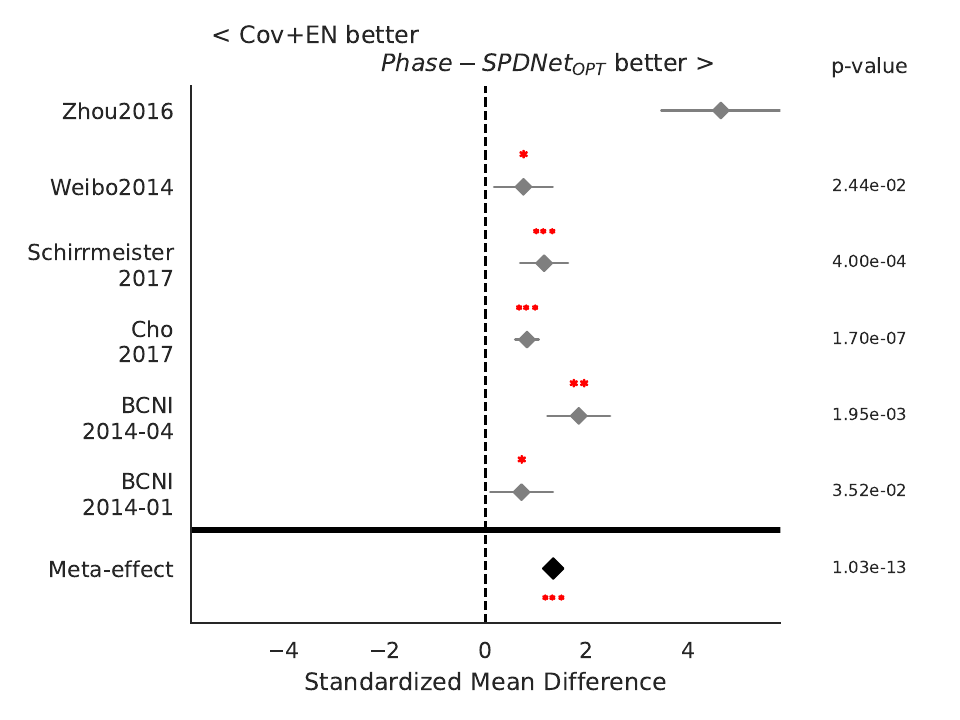}}
            \hfill
   \subfloat[]{%
            \includegraphics[width=0.33\linewidth]{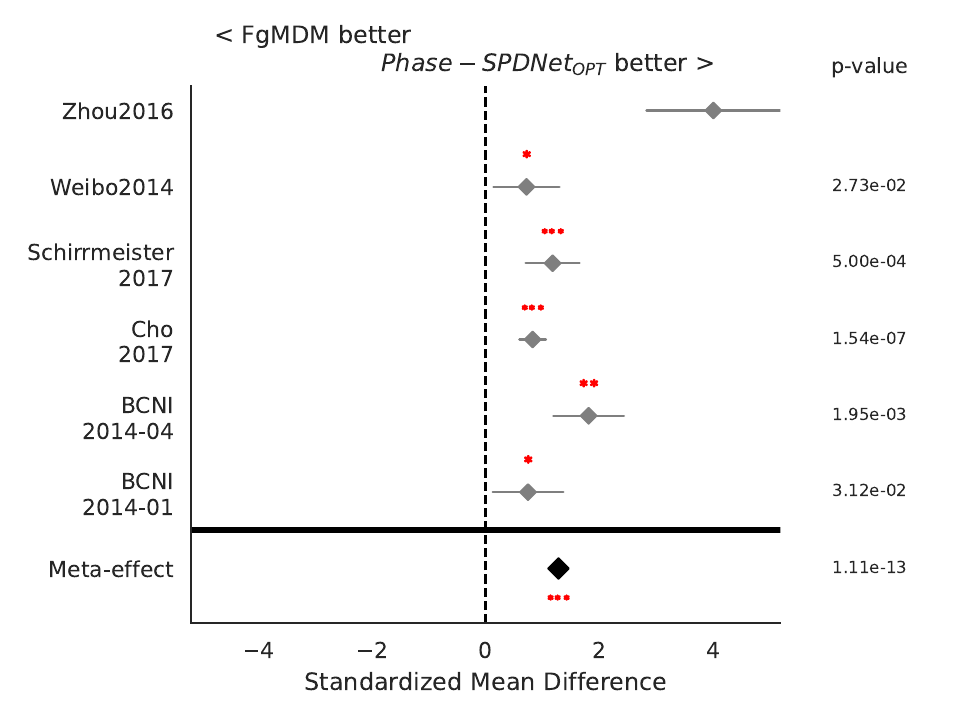}}
            \hfill

    \caption{Result for right hand vs left hand classification, using within-session evaluation for state-of-the-art imaginary coherence pipelines. Plot (a) provides the relative improvement of the method considered with respect to the standard SPDNet of the different pipelines considered. Plot (b) shows a combined meta-analysis (over all datasets) of the different pipelines. It shows the significance of the algorithm on the y-axis being better than the one on the x-axis. The gray level represents the significance level of the ROC-AUC difference in terms of t-values.  We only show significant interactions ($p < 0.05$). Plots (c), (d), and (e) show the meta-analysis of \method$_{OPT}$ against SPDNet, COV+EN, and FgMDM, respectively. We show the standardized mean differences of p-values computed as a one-tailed Wilcoxon signed-rank test for the hypothesis given in the plot title. The {\color{gray}\textbf{gray}} bar denotes the $95\%$ interval. {\color{red}{\textbf{*}}} stands for $p < 0.05$, {\color{red}{\textbf{**}}} for $p < 0.01$, and {\color{red}{\textbf{***}}} for $p < 0.001$.
    }
    \label{fig:rhlh-CohImm}
\end{figure*}

\begin{figure*}[ht]  
    \centering
    \centering
         \subfloat[]{%
            \includegraphics[width=0.65\linewidth]{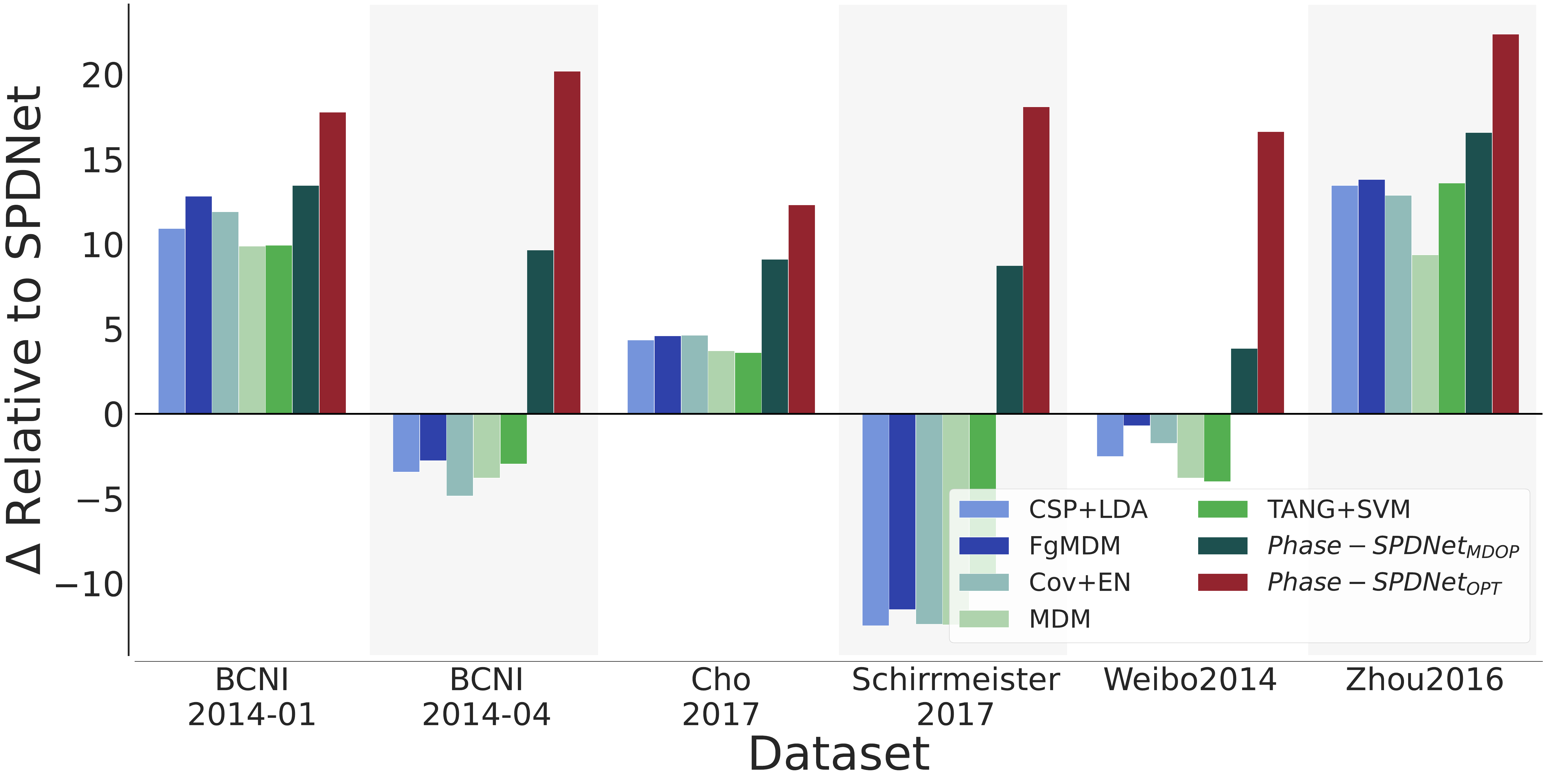}}  
                        \hfill
    \subfloat[]{%
            \includegraphics[width=0.33\linewidth]{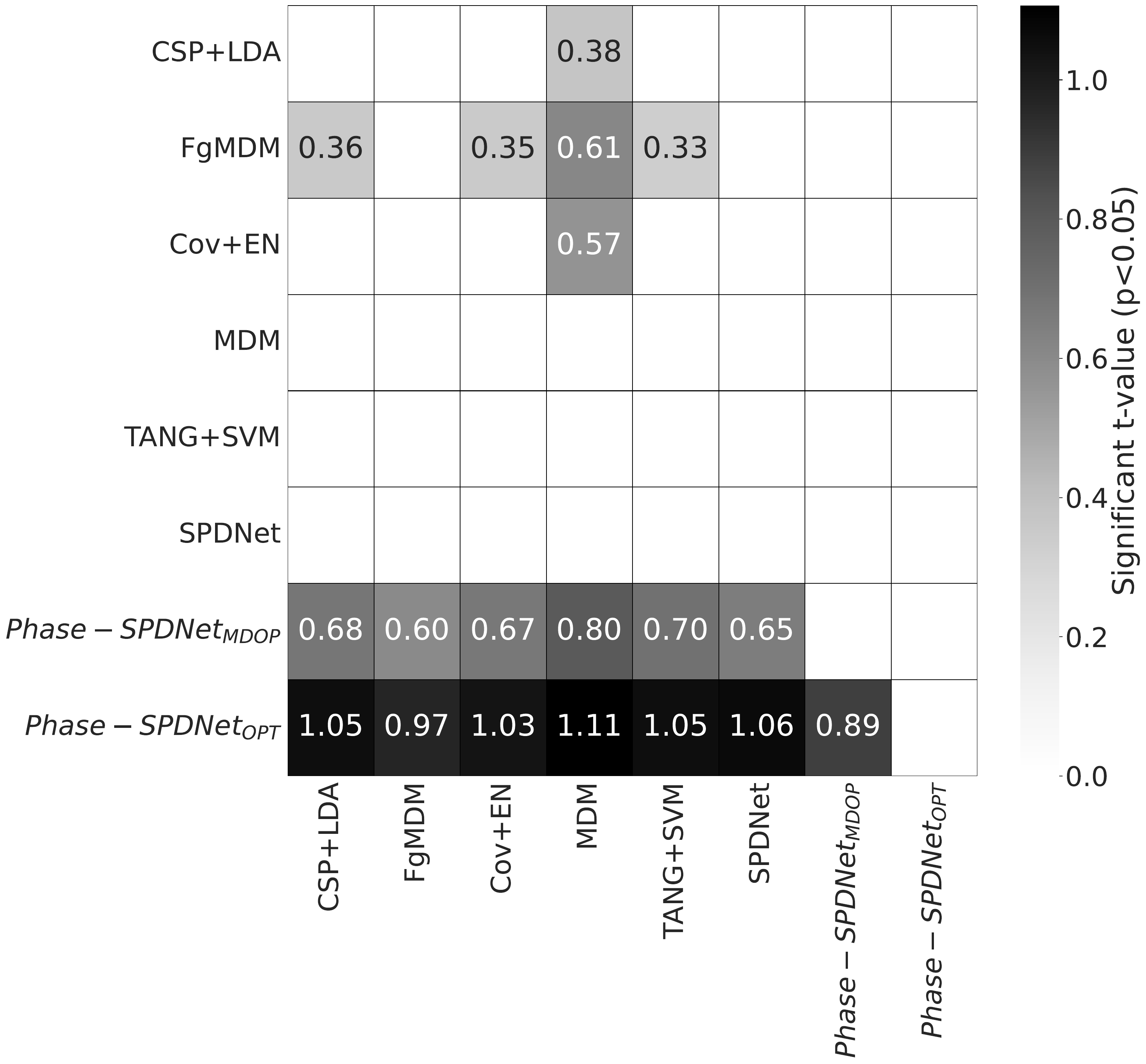}}
    \\
   \subfloat[]{%
            \includegraphics[width=0.33\linewidth]{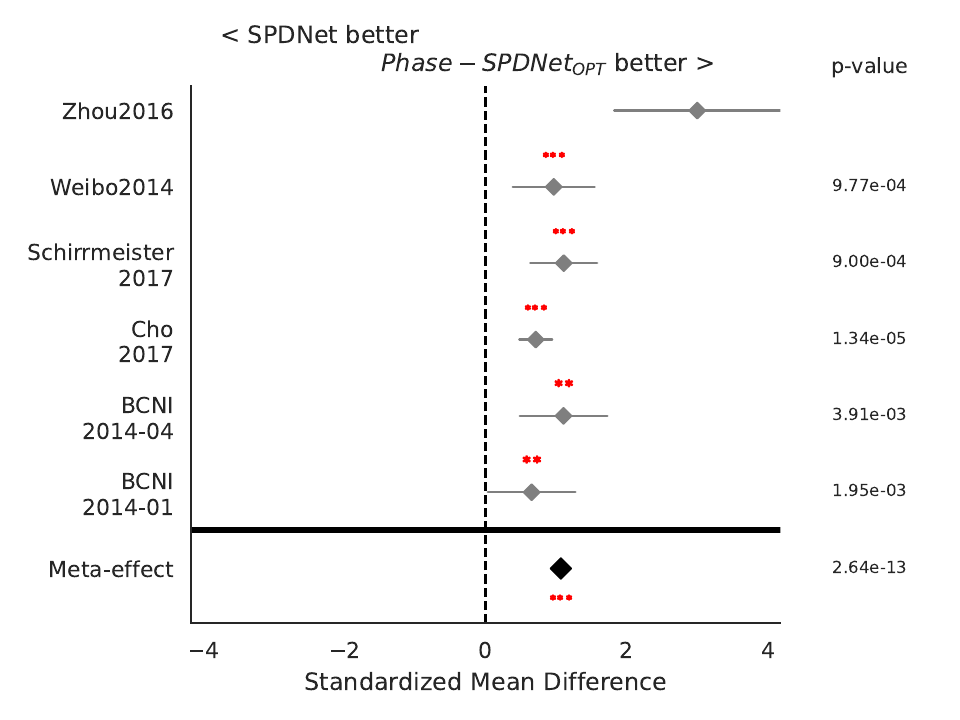}}
            \hfill
   \subfloat[]{%
            \includegraphics[width=0.33\linewidth]{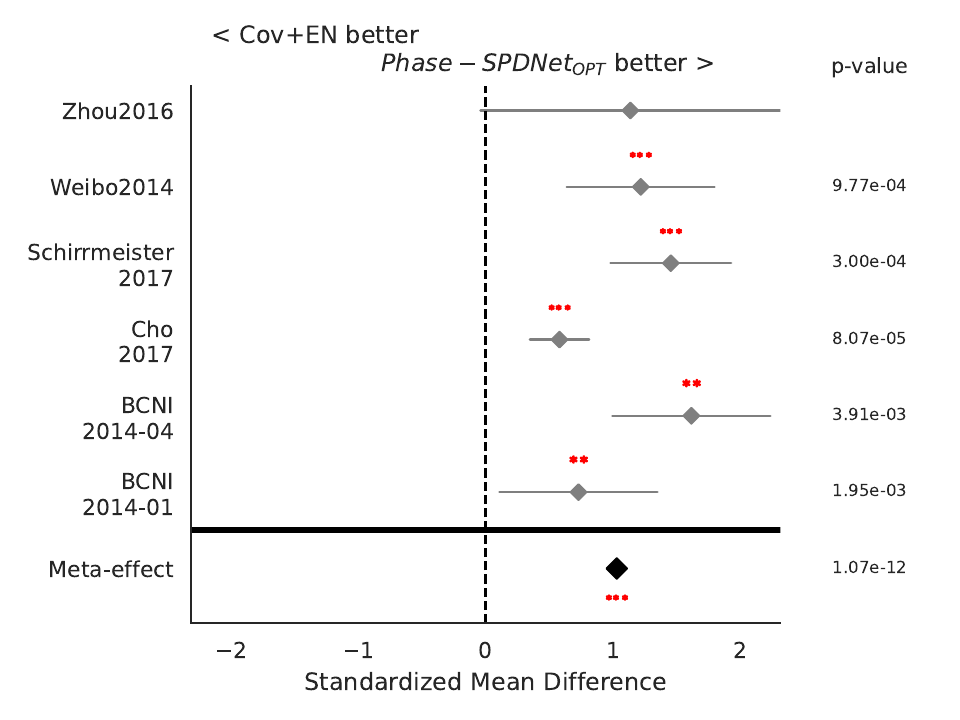}}
            \hfill
   \subfloat[]{%
            \includegraphics[width=0.33\linewidth]{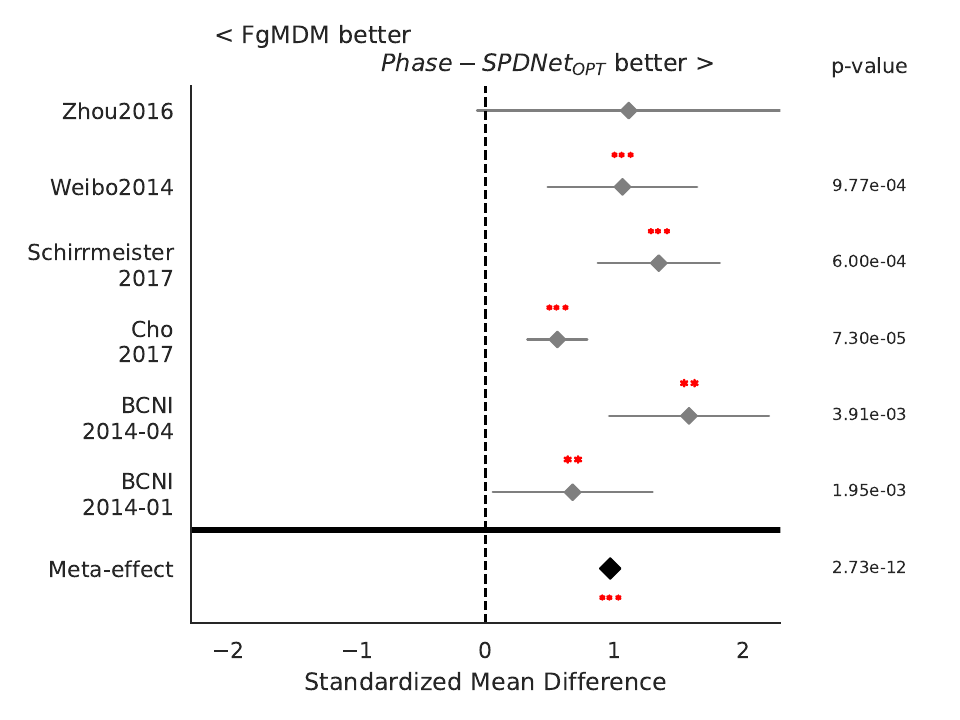}}
            \hfill

    \caption{Result for right hand vs left hand classification, using within-session evaluation for state-of-the-art instantaneous coherence pipelines. Plot (a) provides the relative improvement of the method considered with respect to the standard SPDNet of the different pipelines considered. Plot (b) shows a combined meta-analysis (over all datasets) of the different pipelines. It shows the significance of the algorithm on the y-axis being better than the one on the x-axis. The gray level represents the significance level of the ROC-AUC difference in terms of t-values.  We only show significant interactions ($p < 0.05$). Plots (c), (d), and (e) show the meta-analysis of \method$_{OPT}$ against SPDNet, COV+EN, and FgMDM, respectively. We show the standardized mean differences of p-values computed as a one-tailed Wilcoxon signed-rank test for the hypothesis given in the plot title. The {\color{gray}\textbf{gray}} bar denotes the $95\%$ interval. {\color{red}{\textbf{*}}} stands for $p < 0.05$, {\color{red}{\textbf{**}}} for $p < 0.01$, and {\color{red}{\textbf{***}}} for $p < 0.001$.
    }
    \label{fig:rhlh-CohInsta}
\end{figure*}

Results can be found in Table~\ref{table:rhlh_coh} while the detailed statistical analysis can be found in Figures~\ref{fig:rhlh-CohImm} and \ref{fig:rhlh-CohInsta}.

\begin{table}[!ht]
\caption{Summary of performances via average AUC-ROC (\%) on six datasets (BCNI2014-01, BCNI2014-04, Cho2017, Schirrmeister2017, Weibo2014, and Zhou2016), for a left vs. right motor imagery task. Bold numbers represent the best score in each dataset. Estimators Imaginary Coherence (Imag) and Instantaneous Coherence (Inst)}.\\
\resizebox{1.0\linewidth}{!}{
\begin{tabular}{c|c|c|c|c|c|c|c}
\toprule 

\multirow{1}{*}{\bf Models} & \multirow{1}{*}{\bf Estimator} & \multicolumn{1}{|c|}{\bf BNCI2014001} & \textbf{BNCI2014004} & \textbf{Cho2017} & \textbf{Schirrmeister2017} &               \textbf{Weibo2014} & \textbf{Zhou2016}\\
\midrule
SPDNet & Imag & 56.55$\pm$7.41 & 58.12$\pm$12.20 & 56.16$\pm$8.94 & 59.15$\pm$7.84 & 60.59$\pm$10.78 & 62.89$\pm$8.97\\ 
CSP+LDA & Imag & 57.66$\pm$7.19 & 53.09$\pm$8.55 & 51.63$\pm$6.30 & 53.06$\pm$5.20 & 56.77$\pm$7.62 & 64.47$\pm$7.47\\ 
TANG+SVM & Imag & 59.39$\pm$8.86 & 53.66$\pm$8.26 & 52.45$\pm$6.24 & 52.93$\pm$4.83 & 56.09$\pm$8.89 & 64.19$\pm$9.49\\ 
FgMDM & Imag & 60.19$\pm$7.72 & 52.83$\pm$8.90 & 51.93$\pm$6.39 & 53.99$\pm$4.57 & 57.50$\pm$7.92 & 65.52$\pm$8.72\\ 
MDM & Imag & 60.30$\pm$7.83 & 54.04$\pm$9.00 & 52.09$\pm$6.67 & 53.36$\pm$5.34 & 57.31$\pm$7.33 & 66.27$\pm$8.60\\ 
Cov+EN & Imag & 60.71$\pm$7.92 & 52.67$\pm$8.96 & 52.07$\pm$6.70 & 54.42$\pm$4.07 & 57.21$\pm$8.23 & 64.58$\pm$8.41\\ 
\method$_{MDOP}$ (Our) & Imag & 60.89$\pm$10.17 & 65.70$\pm$14.06 & 56.51$\pm$8.60 & 61.69$\pm$9.81 & 64.75$\pm$11.33 & 72.16$\pm$7.40\\ 
\method$_{OPT}$ (Our) & Imag & \textbf{67.78$\pm$16.44} & \textbf{76.25$\pm$15.91} & \textbf{60.79$\pm$11.57} & \textbf{69.54$\pm$15.44} & \textbf{71.53$\pm$18.80} & \textbf{88.76$\pm$5.57}\\ 
\hline 
\hline SPDNet & Inst & 63.42$\pm$5.76 & 63.71$\pm$12.88 & 57.61$\pm$11.09 & 59.73$\pm$6.89 & 66.11$\pm$12.02 & 75.07$\pm$7.63\\ 
MDM & Inst & 69.74$\pm$14.19 & 59.89$\pm$10.72 & 58.76$\pm$9.65 & 52.01$\pm$5.77 & 63.24$\pm$14.25 & 81.94$\pm$8.91\\ 
TANG+SVM & Inst & 69.82$\pm$14.76 & 60.44$\pm$9.76 & 58.86$\pm$10.11 & 53.31$\pm$5.42 & 63.10$\pm$15.51 & 85.02$\pm$8.29\\ 
CSP+LDA & Inst & 70.49$\pm$14.80 & 60.13$\pm$10.23 & 59.23$\pm$10.40 & 51.94$\pm$6.06 & 64.32$\pm$15.92 & 84.95$\pm$8.94\\ 
Cov+EN & Inst & 71.01$\pm$14.33 & 59.34$\pm$10.89 & 59.37$\pm$9.86 & 52.06$\pm$5.39 & 64.67$\pm$14.79 & 84.53$\pm$8.81\\ 
FgMDM & Inst & 71.61$\pm$13.86 & 60.64$\pm$10.24 & 59.45$\pm$10.29 & 52.53$\pm$5.32 & 65.09$\pm$14.73 & 85.17$\pm$8.36\\ 
\method$_{MDOP}$ (Our) & Inst & 71.84$\pm$16.75 & 68.83$\pm$17.41 & 62.04$\pm$12.02 & 65.37$\pm$12.91 & 68.31$\pm$14.17 & 87.03$\pm$6.13\\ 
\method$_{OPT}$ (Our) & Inst & \textbf{73.08$\pm$17.23} & \textbf{75.95$\pm$17.88} & \textbf{64.13$\pm$12.83} & \textbf{71.15$\pm$15.94} & \textbf{77.01$\pm$17.98} & \textbf{91.14$\pm$5.30}\\ 
 \bottomrule
\end{tabular}}
\label{table:rhlh_coh}
\end{table}

In this study, the situation is even rosier: \method$_{OPT}$, used with either imaginary or instantaneous coherence features, emerges as the best pipeline across all datasets. Interestingly, this study, which focuses on right versus left-hand motor imagery classification, ranks the information content in a ranking where the Imagination-based algorithm outperforms the instantaneous-based one. This result contrasts with the state-of-the-art understanding, which suggests that Instantaneous coherence mitigates the effects of volume conduction~\cite{nolte2004identifying}.

When using imaginary coherence, SPDNet alone already achieves statistically superior results compared to state-of-the-art approaches employing the same features. However, the augmentation procedure introduces a significant performance boost, a finding supported by the meta-effect analysis shown in Fig~\ref{fig:rhlh-CohImm} (c).

In the case of instantaneous coherence, on the other hand, the impact of the augmentation procedure is even more significant. In fact, with this feature, the standard SPDNet fails to exhibit statistically superior performance in comparison to the state-of-the-art pipelines. \method$_{OPT}$, on the other hand, stands out as the top-performing pipeline, showing that the augmentation procedure brings a statistically significant influence on classification outcomes (see Fig~\ref{fig:rhlh-CohInsta} (a)).

\section{Additional Results}

\begin{table}[!ht]
\begin{center}
    \caption{Parameters common to all DL pipelines.}
\resizebox{0.5\linewidth}{!}{\begin{tabular}{c|c}
  \hline
  \textbf{Parameter}  &  \textbf{Value} \\ \hline
    Epoch & 300 \\ \hline
    Batch Size & 64\\ \hline
    Validation Split & 0.1 \\ \hline
    Loss & Sparse Categorical Crossentropy \\ \hline
    Optimizer & Adam \\
    & Learning Rate = 0.001 \\ \hline
    Callbacks ES & Early Stopping \\
    & Patience = 75 \\
    & Monitor = Validation Loss \\ \hline
    Callbacks LR & ReduceLROnPlateau \\
    & Patience = 75 \\
    & Monitor = Validation Loss \\
    & Factor = 0.5 \\ \hline
\end{tabular}
}
\label{table:pipeline_parameter_DL}
\end{center}
\end{table}

\end{document}